\documentclass[12pt]{article}
\usepackage{amssymb,amsmath}
\usepackage{graphicx}
\usepackage{epsfig}
\usepackage{epstopdf}
\usepackage{latexsym}
\usepackage{psfrag}
\usepackage{booktabs}
\usepackage{bbm}

\usepackage[toc]{appendix}

\usepackage{color}
\usepackage{datetime}
\usepackage[
      colorlinks=false,
      linkcolor=darkblue,  
      urlcolor=blue,    
      filecolor=blue,     
      citecolor=red,
linktocpage=true,
      pdfstartview=FitV,
      bookmarksopen=true    
      ]{hyperref}

\ifpdf
\fi

\numberwithin{equation}{section}

\renewcommand{\theequation}{\arabic{section}.\arabic{equation}}

\def\coeff#1#2{\relax{\textstyle {#1 \over #2}}\displaystyle}
\def\ds{\displaystyle}

\def\IC{\mathbb{C}}
\def\IH{\mathbb{H}}
\def\IP{\mathbb{P}}
\def\IR{\mathbb{R}}
\def\ZZ{\mathbb{Z}}

\def\cF{{\cal F}}

\def\cL{{\cal L}}

\def\cN{{\cal N}}

\definecolor{cardinal}{rgb}{0.6,0,0}
\definecolor{darkgreen}{rgb}{0,0.5,0}
\definecolor{golden}{rgb}{0.92, 0.7, 0}
\definecolor{midnight}{rgb}{0, 0, 0.5}
\definecolor{darkblue}{rgb}{0.2, 0, 0.8}

\begin{document}  

\begin{titlepage}
 
\bigskip
\bigskip
\bigskip
\bigskip
\begin{center} 
{\Large \bf   Hair in the Back of a Throat:}
{\Large \bf   Non-Supersymmetric Multi-Center Solutions from K\"ahler Manifolds}

\bigskip
\bigskip 

{\bf Nikolay Bobev${}^{(1)}$, 
Ben Niehoff${}^{(2)}$ and Nicholas P. Warner${}^{(2)}$ \\ }
\bigskip
${}^{(1)}$
Simons Center for Geometry and Physics\\
Stony Brook University\\
Stony Brook, NY 11794, USA\\
\vskip 5mm
${}^{(2)}$
Department of Physics and Astronomy \\
University of Southern California \\
Los Angeles, CA 90089, USA  \\
\bigskip
nbobev@scgp.stonybrook.edu,~bniehoff@usc.edu,~warner@usc.edu  \\
\end{center}

\begin{abstract}

\noindent We find a class of non-supersymmetric multi-center solutions of the STU model of five-dimensional ungauged supergravity. The solutions are determined by a system of linear equations defined on a four-dimensional K\"ahler manifold with vanishing Ricci scalar and a $U(1)$ isometry.   The most general class of such K\"ahler manifolds was studied by LeBrun and they have non-trivial $2$-cycles that can support the topological fluxes characteristic of bubbled geometries.  After imposing an additional $U(1)$ symmetry on the base we find explicit multi-center supergravity solutions. We show that there is an infinite number of regular multi-center solutions with non-trivial topology that are asymptotic to the near-horizon limit of a BMPV black hole. 

\end{abstract}

\end{titlepage}


\tableofcontents

\section{Introduction}

The construction of microstate geometries for black holes and black rings is now a fairly well developed subject.  The defining idea of such geometries is that they have the same asymptotics at infinity as a given black hole or black ring and yet the solutions are completely smooth and, as a consequence, they resolve the black-hole or black-ring singularity into smooth ``bubbled geometries.''     More generally, it is interesting to see what classes of smooth, multi-centered geometries can be inserted into a black-hole, or black-ring  throat since such additional structure may be thought of as ``hair'' on the black object.   Since the throat geometries are typically asymptotic to an anti de Sitter space, such geometric ``hair in the back of a throat'' can be studied quite precisely using holography.    One of the primary purposes in finding  such backgrounds is not only to elucidate the possible microstate geometries but also to obtain a better semi-classical description of black-hole microstates in general.

Most of the progress on microstate geometries has been for BPS solutions  \cite{Bena:2005va,Berglund:2005vb, Bena:2007kg} because such solutions are generically far simpler and are, in fact, governed by a linear system of differential equations \cite{Bena:2004de}.  However, recent work has shown that there are also linear systems of equations that govern very large families  of extremal, non-BPS solutions \cite{Goldstein:2008fq,Bena:2009ev, Bena:2009en,Bena:2009fi,Bena:2009qv, Bobev:2009kn}. These families not only include most of  the known extremal, non-BPS black holes and black rings but also greatly generalize these solutions by including more charges and extending the results to completely new non-BPS, multi-centered black-ring solutions.  The construction of such solutions is a very important first step in understanding the structure and properties of non-BPS solutions in general, particularly since such solutions have played a pivotal role in the analysis of attractor flows \cite{Ferrara:1995ih, Ferrara:1996dd,Ceresole:2007wx, LopesCardoso:2007ky,Andrianopoli:2007gt, Ferrara:2008ap,Bellucci:2008sv}, black-hole bound states,  deconstruction \cite{Denef:2007yt, Gaiotto:2007ag, Gimon:2009gk}, wall crossing and entropy enigmas \cite{Denef:2002ru, Bates:2003vx, Denef:2007vg}.

Ultimately one would like to find the non-BPS analogs of bubbled solutions in which the singularity of the underlying black object has been resolved and the black-hole, or black-ring, throat is rounded off in some form of smooth geometry.  Such solutions have come to be known as {\it microstate geometries} and the requisite geometric transition to the bubbled geometries  that lies at the heart of such objects has been found and extensively studied for BPS solutions \cite{Bena:2005va,Berglund:2005vb, Bena:2007kg}.  However, there are very few examples of the corresponding smooth, non-BPS geometries    \cite{Bena:2009fi, Bena:2009qv, Bobev:2009kn}.  Indeed, there are rather few examples of non-BPS solutions with more than one non-trivial bubble in a black-hole background.  One of the purposes of this paper is to investigate an interesting new class of bubbled geometries and to find new non-BPS bubbled solutions.  

A crucial first step in finding bubbled geometries is to find smooth, four-dimensional ``base geometries'' with non-trivial homology (the ``bubbles'')  that supports the cohomological fluxes that give rise to the charges of the solution.  In BPS solutions (and in some non-BPS solutions) these base geometries are necessarily hyper-K\"ahler manifolds and are usually chosen to be asymptotic to $\IR^4$ or $\IR^3 \times S^1$ so that the final, five-dimensional geometry is asymptotic to five-dimensional, or four-dimensional, Minkowski space.  There are a rich variety of such admissible base geometries because the metrics are allowed to be ambi-polar, that is, the base metric is allowed to reverse sign from regions with signature $+4$ to regions with signature $-4$.  In spite of this apparent peculiarity, the presence of very non-trivial warp factors means that such base metrics can give rise to perfectly viable, smooth, Lorentzian five-dimensional geometries.  The ambi-polar generalizations of Gibbons-Hawking (GH) metrics \cite{Gibbons:1979zt} have proven extremely valuable in that they have given rise to extensive and  highly computable classes of BPS microstate geometries.

It was one of the important realizations of \cite{Bena:2009fi, Bena:2009qv} that one could relax the hyper-K\"ahler condition on the base geometry and still obtain non-BPS solutions from linear systems of equations.  In particular, it was shown in  \cite{Bena:2009fi} how this could be achieved by starting from a four-dimensional Euclidean solution to Einstein's equations coupled to an electromagnetic field.  This led to some interesting examples based upon Israel-Wilson metrics but it turns out that many of these naturally arise as spectral flows of more standard non-BPS solutions based upon Gibbons-Hawking  metrics. One can also construct a regular non-BPS solution with an Euclidean Kerr-Newman base \cite{Bobev:2009kn}, however the base has only one topological two-cycle and there is no obvious way to generalize it easily. A natural follow-up question is whether there are any other classes of Euclidean electrovac solutions that contain appropriate non-trivial homology and that can be used to generate non-BPS microstate geometries.  It turns out that there is a relatively straightforward  generalization of the Gibbons-Hawking metrics to K\"ahler electrovac solutions and these geometries were studied extensively by LeBrun in \cite{LeBrun:1991}.

The central result in  \cite{LeBrun:1991} is to find the explicit local form of all Euclidean, four-dimensional K\"ahler metrics  that have a $U(1)$ isometry and  a vanishing Ricci scalar.  It is then shown in a follow-up paper, \cite{LeBrun:2008kh},  that these solutions are necessarily electrovac solutions whose electromagnetic field is related to the K\"ahler form.  This generalizes earlier work on the classification of  hyper-K\"ahler metrics with $U(1)$ isometry \cite{Boyer:1982mm,DasGegenberg} in which it is shown that the metric is determined by a single function that is required to satisfy the Affine Toda equation.  The Gibbons-Hawking metrics then emerge as precisely the metrics for which the $U(1)$ action preserves all three complex structures (that is, the $U(1)$ is tri-holomorphic).  The ``LeBrun metrics'' in  \cite{LeBrun:1991} are defined by two functions, one of which must satisfy the Affine Toda equation and the other of which must essentially be harmonic in a background defined by the Affine Toda solution.   This family of solutions collapses to either the general class of $U(1)$-invariant  hyper-K\"ahler metrics, or to the GH family if one makes an essentially trivial choice for one of the two functions.

Amongst the general class of LeBrun metrics is what we will call the ``LeBrun-Burns metrics,''  which provide a simple, explicit class of K\"ahler metrics on $\IC^2$ with $n$ points blown up\footnote{One of the  original motivations in \cite{LeBrun:1991} for finding this metric is that there is a simple conformal compactification that yields an explicit K\"ahler metric on the connected sum of several $\IC \IP_2$'s.}.  The end result is rather similar to the Gibbons-Hawking metrics except that the $\IR^3$ sections and the harmonic functions on $\IR^3$ are now replaced by the hyperbolic space, $\mathbb{H}^3$, and its harmonic functions.   The LeBrun-Burns metrics are also asymptotic to $\IR^4$ and, interestingly enough, the $U(1)$ isometry on this  $\IR^4$ does {\it not} act in a manner that matches the  tri-holomorphic $U(1)$ action on a Gibbons-Hawking metric that is similarly  asymptotic to $\IR^4$.    Thus the solutions obtained from the LeBrun-Burns metrics will be intrinsically different from those obtained from GH solutions.  

In this paper we explicitly solve the equations of motion of five-dimensional supergravity on an axisymmetric LeBrun-Burns base and thus provide an infinite class of non-supersymmetric multi-centered solution. We find that, due to the Maxwell flux on the four-dimensional K\"ahler base, the five-dimensional backgrounds are not asymptotically flat and have the asymptotics of a warped, rotating $AdS_2\times S^3$ space. For certain choices of parameters this becomes the near horizon limit of a BMPV black hole \cite{Breckenridge:1996is}.  We would like to emphasize that although our approach to finding these solutions was inspired by the construction of BPS microstate geometries we find the most general axisymmetric solutions within the floating brane Ansatz  \cite{Bena:2009fi} and with an axisymmetric LeBrun-Burns base. In particular our solutions include superpositions of non-supersymmetric concentric black rings and other potentially interesting solutions with horizons.

In section 2 of this paper we review the process through which one can construct five-dimensional non-BPS solutions  using the floating-brane Ansatz \cite{Bena:2009fi}. In Section 3 we introduce the general LeBrun metrics and begin solving the linear system in this background, and in Section 4 we specialize to the LeBrun-Burns metrics, discuss their properties and further reduce the linear system and exhibit all the necessary Green functions. We will also show, in Section 4,  that because the Maxwell field on the four-dimensional base involves the K\"ahler form, the energy-momentum tensor does not fall off at infinity and hence the LeBrun-Burns base metrics do not naturally lead to five-dimensional solutions that are asymptotic to flat space.   

In Section 5 we start with the simplest possible LeBrun-Burns metric, $\mathbb{R}^4$, and use the linear system and Green functions of Section 4 to construct very simple, explicit examples of non-BPS solutions. We find that the natural asymptotic geometries that arise from LeBrun-Burns base metrics are a warped, rotating $AdS_2\times S^3$ space.    In Section 6 we consider more general LeBrun-Burns base metrics with non-trivial topology and find the general axisymmetric solution to the system of non-BPS equations on such a base.  Five-dimensional regularity and the absence of closed time-like curves (CTCs) puts very stringent conditions on our solutions but in spite of this we find a family of regular solutions with non-trivial bubbles that are asymptotic to the near-horizon region of a BMPV black hole. In Section 7 we present our conclusions and suggest some directions for further work. Some of the technical details of the construction of the solutions are relegated to the Appendix.

\section{The  family of non-BPS solutions}

It is simplest to characterize our solutions in terms of $\cN \! = \!  2$, five-dimensional ungauged supergravity with three $U(1)$ gauge fields.  This theory may also be thought of as  arising from a truncation of eleven-dimensional supergravity on $T^6$.  The bosonic action is given by:
\begin{eqnarray}
  S = \frac {1}{ 2 \kappa_{5}} \int\!\sqrt{-g}\,d^5x \Big( R  -\coeff{1}{2} Q_{IJ} F_{\mu \nu}^I   F^{J \mu \nu} - Q_{IJ} \partial_\mu X^I  \partial^\mu X^J -\coeff {1}{24} C_{IJK} F^I_{ \mu \nu} F^J_{\rho\sigma} A^K_{\lambda} \bar\epsilon^{\mu\nu\rho\sigma\lambda}\Big) \,,
  \label{5daction}
\end{eqnarray}
where we use the conventions of \cite{Bena:2009fi}.   The matrix that defines the kinetic terms can be written as:
\begin{equation}
  Q_{IJ} ~=~    \frac{1}{2} \,{\rm diag}\,\big((X^1)^{-2} , (X^2)^{-2},(X^3)^{-2} \big) \,.
\label{scalarkinterm}
\end{equation}
The scalar fields themselves are not independent and are most conveniently parametrized in terms of three other scalar fields, $Z_I$:  
\begin{equation}
X^1    =\bigg( \frac{Z_2 \, Z_3}{Z_1^2} \bigg)^{1/3} \,, \quad X^2    = \bigg( \frac{Z_1 \, Z_3}{Z_2^2} \bigg)^{1/3} \,,\quad X^3   =\bigg( \frac{Z_1 \, Z_2}{Z_3^2} \bigg)^{1/3}  \,,
\label{XZrelns}
\end{equation}
which satisfy the constraint $X^1 X^2 X^3 =1$.    The reason why we use three scalars, $Z_I$, to parametrize two independent scalar fields  becomes evident once we write the metric Ansatz
\begin{equation}
ds_5^2 ~=~ -Z^{-2} \,(dt + k)^2 ~+~ Z \, ds_4^2  \,,
\label{metAnsatz}
\end{equation}
and define $Z$ by:
\begin{equation}
Z ~\equiv~ \big( Z_1 \, Z_2 \, Z_3  \big)^{1/3}   \,.
\label{Zdefn}
\end{equation}
This turns out to be an extremely convenient way to express both the scalar fields and the warp factors.

The electromagnetic fields are given by the  ``floating brane'' Ansatz of \cite{Bena:2009fi}, which relates metric coefficient and scalar fields  to the electrostatic potentials.  In particular, the Maxwell potentials are given by:
\begin{equation}
A^{(I)}   ~=~  - Z_I^{-1}\, (dt +k) + B^{(I)}  \,,
\label{AAnsatz}
\end{equation}
where $B^{(I)}$ is a one-form on the base $ds_4^2$.   It is convenient to introduce the magnetic two-from field strengths
\begin{equation}
	\Theta^{(I)} ~\equiv~ d B^{(I)}\,.
\end{equation}

The four-dimensional base space, $ds^2_4$, has to be a solution of Euclidean Einstein-Maxwell equations\footnote{The normalization of the fields in this equation is different from most standard sources on general relativity and is chosen to agree with the four-dimensional conventions in \cite{Bena:2009fi}.} 
\begin{equation}
{R}_{\mu\nu} = \coeff{1}{2}\, \big( \cF_{\mu\rho} {\cF_{\nu}}^{\rho} -  \coeff{1}{4}\, g_{\mu\nu} \cF_{\rho\sigma} \cF^{\rho\sigma} \big)\,,
 \label{electrovacequation}
\end{equation}
where all quantities are computed in the four-dimensional base metric. The Maxwell field is then decomposed as:
\begin{equation} \label{Fdecomp}
\cF = \Theta^{(3)} - \omega^{(3)}_{-}\,.
\end{equation}
where $\Theta^{(3)}$ is self-dual and $\omega^{(3)}_{-}$ is anti-self-dual.  The equations of motion $d \cF = d * \cF =0$ imply that $\Theta^{(3)}$ and $\omega^{(3)}_{-}$ are harmonic.  As the notation implies, this defines the magnetic two-from field strength $\Theta^{(3)}$.

The linear system that solves the equations of motion can now be written as \cite{Bena:2009fi}:   
\begin{eqnarray}
 \hat \nabla^2 Z_1 &=&   *_4  \big[ \Theta^{(2)} \wedge \Theta^{(3)}   \big]    \,,  \qquad \big(\Theta^{(2)} -    *_4 \Theta^{(2)} \big)  ~=~  2  \, Z_1 \, \omega^{(3)}_- \,,  \label{ZMax1} \\
\hat \nabla^2 Z_2&=& *_4  \big[ \Theta^{(1)} \wedge \Theta^{(3)}   \big]  \,,  \qquad
 \big(\Theta^{(1)} -   *_4 \Theta^{(1)} \big)    ~=~ 2 \, Z_2 \, \omega^{(3)}_-  \label{ZMax2}\,,
\end{eqnarray}
and
\begin{eqnarray}
 \hat \nabla^2 Z_3  &=&  *_4  \big[ \Theta^{(1)} \wedge \Theta^{(2)}   ~-~    \omega^{(3)}_- \wedge(  dk   -  *_4  dk  )  \big]    \,, 
\label{Z3eqn} \\
 dk    ~+~   *_4 dk    &=&  \frac{1}{2} \, \sum_I \, Z_I \,  \big(\Theta^{(I)} +  *_4 \Theta^{(I)} \big)     \,,  \label{keqn} 
\end{eqnarray}
where $\hat \nabla^2$ is the Laplacian on $ds^2_4$. Indeed, having chosen the electrovac solution that defines the base metric, one uses \eqref{Fdecomp} to read off $\Theta^{(3)}$ and $\omega_-^{(3)}$.  As a result,    \eqref{ZMax1} and \eqref{ZMax2} are two linear coupled equations for $Z_1$ and $\Theta^{(2)}$ and $Z_2$ and $\Theta^{(1)}$ respectively.  Once these equations are solved, $k$ and $Z_3$ are solutions to the system of linear equations \eqref{Z3eqn} and \eqref{keqn}. Our purpose here is to implement this procedure for the  LeBrun metrics.

\section{The non-BPS equations for the LeBrun metrics}

\subsection{The metric}
\label{LeBrunMet}

The LeBrun metric, \cite{LeBrun:1991}, is
\begin{equation}
ds^2_{4}~=~  w^{-1}\, (d\tau + A)^2 ~+~  w\, (e^{u} (dx^2+dy^2) +dz^2) \,,
\label{LBmet}
\end{equation}
where $u$ and $w$ are two functions of $(x,y,z)$ which obey the $su(\infty)$ Toda equation and its linearized form:
\begin{eqnarray}
&&\partial_{x}^2 \,u ~+~ \partial_{y}^2 \,u ~+~ \partial_z^2\, (e^u) ~=~  0 \,,  \label{Toda} \\
&&\partial_{x}^2 \,w ~+~ \partial_{y}^2 \,w ~+~ \partial_z^2\,(e^u\, w) ~=~ 0\,.\label{linearToda}
\end{eqnarray}
The one-form, $A$, satisfies:  
\begin{equation}
dA ~=~ \partial_{x}w ~dy\wedge dz ~-~  \partial_{y}w~ dx\wedge dz ~+~ \partial_z(e^u w) ~dx \wedge dy\,, \label{Adefn}
\end{equation}
and the integrability of this differential, $d^2 A=0$, is equivalent to the equation (\ref{linearToda}).

The metric is K\"ahler, with K\"ahler form:
\begin{equation}
J ~=~ (d\tau + A) \wedge dz ~-~   w\, e^u \,  dx\wedge dy  \,. \label{Kform}
\end{equation}

It is convenient to introduce frames:
\begin{equation}
 e^0 ~\equiv~  w^{-{1 \over 2}} \,  (d\tau + A)   \,,  \qquad  e^1 ~\equiv~  w^{ {1 \over 2}} \,  e^{ {u \over 2}} \, dx   \,,   
\qquad e^2 ~\equiv~   w^{ {1 \over 2}} \,  e^{ {u \over 2}} \, dy  \,,  \qquad e^3 ~\equiv~  w^{ {1 \over 2}} \,  dz  \,,  \label{frames}
\end{equation}
and the self-dual forms
\begin{eqnarray}
&& \Omega^{(1)}_+  ~\equiv~ e^{- {u \over 2}}  ( e^0 \wedge e^1 ~+~ e^2 \wedge e^3) ~=~ (d\tau + A) \wedge dx ~+~   w\,   dy \wedge dz \,,  \nonumber   \\
&&  \Omega^{(2)}_+  ~\equiv~ e^{- {u \over 2}}  ( e^0 \wedge e^2 ~-~ e^1 \wedge e^3) ~=~ (d\tau + A) \wedge dy ~-~   w\,   dx \wedge dz \,,    \label{sdforms}\\
&&  \Omega^{(3)}_+  ~\equiv~   ( e^0 \wedge e^3 ~+~ e^1 \wedge e^2) ~=~ (d\tau + A) \wedge dz ~+~   w\,  e^u\, dx \wedge dy \,.  \nonumber 
\end{eqnarray}
We will also frequently denote the coordinates by $\vec y \equiv (y_1, y_2, y_3) =(x,y,z) $.

\subsection{Harmonic fluxes}
\label{harmflux}

One can then verify that:
\begin{equation}
\Theta ~\equiv~  \sum_{a=1}^3 \bigg(  \partial_a \bigg( {H \over w} \bigg)\bigg) \,   \Omega^{(a)}_+ \label{harmform1}
\end{equation}
is harmonic if and  only if $H$ satisfies (\ref{linearToda}):
\begin{equation}
\partial_{x}^2\, H ~+~ \partial_{y}^2\, H ~+~ \partial_z^2\, (e^u  \, H ) ~=~  0 \,. \label{harm1}
\end{equation}
Note that if one differentiates (\ref{Toda}) with respect to $z$ one finds that $H = \partial_z u$ satisfies (\ref{harm1}) and (\ref{Toda}).

Define the Maxwell field, $\cF$, by 
\begin{equation}
\cF  ~\equiv~  \Theta ~+~ \alpha \, J  \,, \quad {\rm with}  \quad H ~=~ -{1 \over 2\, \alpha} \, \partial_z u \,. \label{MaxF}
\end{equation}
The metric, (\ref{LBmet}), is then a solution to the Einstein-Maxwell equations (\ref{electrovacequation}).

This fits the form of the linear system obtained in \cite{Bena:2009fi} and so we take:
\begin{equation}
\Theta^{(3)}  ~=~  -{1 \over 2\, \alpha} \,   \sum_{a=1}^3 \bigg( \partial_a \bigg( {\partial_z u \over w} \bigg)\bigg) \,  \Omega^{(a)}_+ \,, \qquad  \omega^{(3)}_- ~=~ - \alpha \, J \,.  \label{Max3}
\end{equation}

One can now use this in the linear system  \eqref{ZMax1},  \eqref{ZMax2},  \eqref{Z3eqn} and \eqref{keqn} to solve the equations of motion of supergravity.  Note that we can absorb the constant $\alpha$ by rescaling the coordinate $\tau$ and the function $w$ (which in turn rescales the one-form, $A$).  Throughout the rest of this paper, we set $\alpha = -1$.

\subsection{Solving the first layer}
\label{layer1}

One finds that the first part of the system is solved by an Ansatz:
\begin{equation}
\Theta^{(1)}~\equiv~ Z_2 \, J ~+~   \sum_{a=1}^3 p^{(1)}_a \,   \Omega^{(a)}_+ \,, \qquad
\Theta^{(2)}~\equiv~ Z_1 \, J ~+~   \sum_{a=1}^3 p^{(2)}_a \,   \Omega^{(a)}_+   \,, \label{Thetaforms}
\end{equation}
\begin{equation}
Z_1 ~=~ \frac{1}{2}\,   \Big(  {K^{(2)} \, \partial_z u \over w} \Big) ~+~L_1\,, \qquad Z_2 ~=~  \frac{1}{2}\,   \Big( { K^{(1)} \, \partial_z u \over w} \Big) ~+~L_2   \,. \label{Z12form}
\end{equation}
The Bianchi identities imply the following relations
\begin{eqnarray}
&& p^{(1)}_1 ~=~  \partial_x \Big( {K^{(1)} \over w} \Big)  \,, \qquad  p^{(1)}_2 ~=~  \partial_y \Big( {K^{(1)} \over w} \Big)  \,, \qquad p^{(1)}_3 ~=~   -\, Z_2  ~+~ \partial_z \Big( {K^{(1)} \over w} \Big)   \,, \label{pfns1}  \\
&&p^{(2)}_1 ~=~  \partial_x \Big( {K^{(2)} \over w} \Big)  \,, \qquad  p^{(2)}_2 ~=~  \partial_y \Big( {K^{(2)} \over w} \Big)  \,, \qquad p^{(2)}_3 ~=~   -\, Z_1  ~+~ \partial_z \Big( {K^{(2)} \over w} \Big)   \,.  \label{pfns2} 
\end{eqnarray}
One  can add arbitrary functions of $z$ alone to the $p^{(I)}_3$, but these can be absorbed by a shift $K^{(I)} \to K^{(I)} + w \, g^{(I)}(z)$, for some function $g^{(I)}$.

The functions $L_1$ and $L_2$ can be any solution of (\ref{linearToda}), that is:
\begin{equation}
\partial_{x}^2 \,L_I ~+~ \partial_{y}^2 \,L_I ~+~ \partial_z^2\,(e^u\, L_I ) ~=~ 0\,,   \qquad I =1,2\,,\label{Leqns1}
\end{equation}
 and given these solutions the functions $K^{(1)}$ and $K^{(2)}$ are determined by the linear equations:
\begin{eqnarray}
&&\partial_{x}^2 \, K^{(1)}  ~+~ \partial_{y}^2 \,K^{(1)}  ~+~ \partial_z \,(e^u\, \partial_z \,K^{(1)} ) ~=~  2\,  \partial_z \,(e^u\, w\,   L_2 )   \,, \label{lin2a}  \\
&&\partial_{x}^2 \,K^{(2)}  ~+~ \partial_{y}^2 \,K^{(2)}  ~+~ \partial_z \,(e^u\, \partial_z \,K^{(2)} ) ~=~  2\,  \partial_z \,(e^u\, w\,   L_1 )   \,. \label{lin2b} 
\end{eqnarray}
%

\subsection{Solving the second layer}
\label{layer2}

As usual one makes the Ansatz
\begin{equation}
k ~\equiv~ \mu \, (d \tau + A)  ~+~    \omega \,,   \label{kansatz}
\end{equation}
where $\omega = \vec\omega \cdot d \vec y$ is a one-form on the three-dimensional base.
One then finds that the solution may be written as: 
\begin{equation}
Z_3 ~=~  \Big(  {K^{(1)} \,  K^{(2)}  \over w} \Big) ~+~L_3 \,,  \label{Z3form}
\end{equation}
\begin{equation}
\mu ~=~   -\frac{1}{2}\, \Big(  {K^{(1)} \,  K^{(2)} \, \partial_z u  \over w^2} \Big) - \frac{1}{2}\, \Big(  {K^{(1)} \,  L_1  + K^{(2)} \,  L_2   \over w} \Big) ~-~ \frac{1}{4}\, \Big(  {\partial_z u \, L_3  \over w} \Big) ~+~ M \,.  \label{muform}
\end{equation}

The functions $L_3$ and $M$ must then satisfy the linear equations:
\begin{eqnarray}
&&\partial_{x}^2 \, M + \partial_{y}^2 \,  M  + \partial_z \,(e^u\, \partial_z \, M ) =   \partial_z \,(e^u \,  L_1 \, L_2 )   \,,  \label{lin3a}  \\
&& \partial_{x}^2 \, L_3  +  \partial_{y}^2 \, L_3  +  e^u\, \partial_z^2 \, L_3 =    -2 \,  e^u \big[  2\, w \,(- L_1 L_2 + \partial_z M) + L_1 \, \partial_z K^{(1)} + L_2 \, \partial_z K^{(2)}  \big]  \,.  \label{lin3b} 
\end{eqnarray}

Finally, the components of $\vec \omega$ are determined from:
\begin{eqnarray}
(\partial_y \, \omega_z -  \partial_z \, \omega_y ) ~+~ (M \partial_x w - w \partial_x M)  &+&\frac{1}{2} \, \sum_{I=1}^2 (K^{(I)} \partial_x L_I -  L_I  \partial_x K^{(I)}) \notag\\ 
 &+& \frac{1}{4} \, \big((\partial_z u) \,  \partial_x L_3 - L_3 \partial_x (\partial_z u)  \big) ~=~ 0 \,,   \label{omx}  \\
- (\partial_x \, \omega_z -  \partial_z \, \omega_x ) ~+~ (M \partial_y w - w \partial_y M)  &+&\frac{1}{2} \, \sum_{I=1}^2 (K^{(I)} \partial_y L_I -  L_I  \partial_y K^{(I)}) \notag\\ 
 &+& \frac{1}{4} \, \big((\partial_z u) \,  \partial_y L_3 - L_3 \partial_y (\partial_z u)  \big) ~=~ 0  \,,  \label{omy} 
 \end{eqnarray}
 \begin{eqnarray}
 (\partial_x \, \omega_y -  \partial_y \, \omega_x ) &+& (M \partial_z (e^u \, w) - e^u \, w  \,  \partial_z M)  ~+~ \frac{1}{2} \, \sum_{I=1}^2 (K^{(I)} \partial_z (e^u \, L_I)  -   e^u \, L_I   \partial_z K^{(I)}) \notag\\ 
 &+& \frac{1}{4} \, \big((\partial_z e^u) \,  \partial_z L_3 - L_3 \partial_z^2 (e^u)  \big)~+~ 2\, e^u \, w\, L_1 \,L_2 ~=~ 0  \,.  \label{omz} 
\end{eqnarray}
The integrability of these equations for $\vec \omega$ follows from the differential equations satisfied by all the other functions.

\subsection{The complete solution}
\label{summary1}

The complete solution is thus obtained by first choosing a background metric, (\ref{LBmet}), with functions $u$ and $w$ that satisfy (\ref{Toda}) and (\ref{linearToda}).  The choice of background fixes one of the three electromagnetic fields, via  (\ref{Kform})and (\ref{MaxF}) and, in particular, determines $\Theta^{(3)}$ via (\ref{Max3}).  One then finds $L_1$ and $L_2$ as  solutions to the homogeneous equations (\ref{Leqns1}) and uses these solutions as sources  in the linear equations, (\ref{lin2a}) and  (\ref{lin2b}), for $K^{(1)}$ and $K^{(2)}$.  These functions then determine the remaining magnetic fluxes,  $\Theta^{(j)}$, and the electrostatic potentials, or warp factors, $Z_j$, $j=1,2$ from (\ref{Thetaforms}), (\ref{Z12form}), (\ref{pfns1}) and (\ref{pfns2}). Next one solves the linear equation, (\ref{lin3a}), for $M$ with a source determined by $L_1 L_2$ and then solves the linear equation for $L_3$, (\ref{lin3b}), whose source is made from $L_j,  K^{(j)}$ and $M$.    The last step is to use all of these functions to solve the linear, first order system, (\ref{omx})--(\ref{omz}) for $\vec \omega$ and obtain $Z_3$ and the angular momentum vector from (\ref{Z3form}), (\ref{muform}) and  (\ref{kansatz}).  While complicated, this is a linear system of equations once one has chosen a solution for the metric function, $u$.

Before concluding, it is worth noting that taking the function, $u$, to be a constant is a trivial solution to (\ref{Toda})  and then   (\ref{linearToda}) becomes the harmonic equation on $\IR^3$ and the metric reduces to the familiar class of Gibbons-Hawking metrics.  Similary, if one takes $w = \partial_z u$ then it also satisfies   (\ref{linearToda}), as can be seen by differentiating (\ref{Toda})  with respect to $z$.  This yields the general class of hyper-K\"ahler metrics with a non-triholomorphic $U(1)$ isometry \cite{Boyer:1982mm,DasGegenberg,Bena:2007ju}, which are based upon the Affine Toda equation.   The resulting system of equations for non-BPS solutions in five dimensions does not, however, reduce to that considered in the bubbling BPS solutions \cite{Bena:2005va,Berglund:2005vb,Bena:2007kg} but is rather more akin to the fluxes considered in \cite{Bena:2009fi}.  This is because the flux background is a mixture of self-dual and anti-self-dual fluxes and these break supersymmetry.  In particular, the anti-self-dual flux is non-normalizable since it is proportional to the complex structure. Thus even the simple Gibbons-Hawking and Toda limits of the LeBrun backgrounds extend the class of solutions considered thus far.  

The LeBrun solutions are four-dimensional Euclidean Einstein-Maxwell solutions and it is natural to ask whether some of them preserve supersymmetry. Supersymmetric solutions of four-dimensional Euclidean Einstein-Maxwell theory were classified in \cite{Gutowski:2010zs}. The maximally supersymmetric solutions are $\mathbb{R}^4$ or $H_2\times S^2$. There are two classes of solutions which preserve half of the supersymmetries - the well-known Gibbons-Hawking solutions and the Euclidean Israel-Wilson metrics discussed in \cite{Dunajski:2006vs}. Therefore the classification of \cite{Gutowski:2010zs} also demonstrates that the general LeBrun solutions, albeit K\"ahler, are non-supersymmetric solutions of Einstein-Maxwell theory.

\section{The LeBrun-Burns metrics}
\label{BurnsSec}

The LeBrun-Burns metrics represent a very natural generalization of the Gibbons-Hawking metrics in that they are four-dimensional K\"ahler metrics with $n$  2-cycles and associated moduli and they satisfy Einstein's equations coupled to a $U(1)$ gauge field.  They were constructed \cite{LeBrun:1991} (see also [3] as cited in \cite{LeBrun:1991})  as explicit metrics on a decompactification of $n\IC\IP_2$, the connected sum of $n$ $\IC\IP_2$'s.  

\subsection{Defining the metric}
\label{LBBDef}
 
The simplest class of metrics arises if one takes:
\begin{equation}
u  ~=~   \log (2 \, z)\,,  \label{LBBuform}
\end{equation}
and it is then convenient to reparametrize by defining:
\begin{equation}
z  ~\equiv~   \coeff{1}{2}\, \zeta^2 \,, \qquad V ~\equiv~ e^u \, w ~=~  2\,  z\, w    ~=~   \zeta^2 \, w \,.  \label{zetaVdefn}
\end{equation}
 The LeBrun-Burns metric can then be written as
\begin{equation}
ds^2_4  ~=~  \zeta^2 \Big[  V^{-1} \, (d\tau+A)^2 + V \Big(  \frac{dx^2 + dy^2 + d \zeta ^2}{\zeta ^2} \Big) \Big]  \label{LBB2}\,.
\end{equation}
Note that the three-dimensional metric is the standard constant-curvature metric on the hyperbolic plane, $\IH^3$: 
\begin{equation}
ds^2_{\IH^3} ~=~  \ds\frac{dx^2 + dy^2 + d\zeta^2}{\zeta ^2} \,. \label{HypMet}
\end{equation}
The equations (\ref{linearToda}) and  (\ref{Adefn}) that define the four-dimensional base imply that $V$ is a harmonic function on the hyperbolic plane and that $A$ is an appropriate one-form on $\IH^3$: 
\begin{equation}
\nabla^2_{\IH^3} V ~=~  0\,,  \qquad\qquad dA ~=~ *_{\IH^3} dV \,. \label{Hypconds}
\end{equation}
%

\subsection{Geometry of the LeBrun-Burns metric}
\label{LBBGeom}

\subsubsection{Asymptotics}
\label{LeBrunAsym}

To avoid a conical singularity at $\zeta =0$, one must have $V \to 1$ at this point so that the metric in the $(\zeta, \tau)$ direction limits to that of $\IR^2$ in polar coordinates.  Thus the metric in the neighborhood of  $\zeta =0$ is that of $\IR^4$ and regularity requires that one restrict the space to $\zeta \ge 0$.  Similarly, if one requires $V \to 1$ at infinity, the space is asymptotic to $\IR^4 = \IC^2$.  Note that the circle defined by $\tau$ lies in an $\IR^2$ plane of the  $\IR^4$, and the associated isometry therefore only commutes with another $U(1)$ factor in the generic $SO(4)$ holonomy of the base metric.    This is quite different from the way in which the isometry associated with the $U(1)$ fiber behaves in GH spaces.

The Green functions of the Laplacian on $\IH^3$ are the functions:
\begin{eqnarray}
G(x,y,\zeta; a,b,c) &\equiv&    \bigg(  \frac{(x-a)^2+(y-b)^2+ \zeta^2+ c^2} { \sqrt{((x-a)^2+(y-b)^2+ \zeta ^2+ c^2)^2 -  4\, c^2\, \zeta^2 }}~-~1 \bigg) \nonumber\\ 
&=&    \bigg(  \frac{(x-a)^2+(y-b)^2+ \zeta^2+ c^2} { \sqrt{((x-a)^2+(y-b)^2+ (\zeta-c)^2) ((x-a)^2+(y-b)^2+ (\zeta+c)^2) }}~-~1 \bigg) \,, \nonumber\\
\label{Gdefn}
\end{eqnarray}
where one should remember that $\zeta \ge 0$ on $\IH^3$ and so this function only has one singularity in the domain of definition.
The constant has been added so that $G$ vanishes at infinity.  Given $G$, we can then solve for $A$ in \eqref{Hypconds}.  Putting $A = D(x,y,\zeta; a,b,c) \; d\phi$, we obtain
\begin{eqnarray}
D(x,y,\zeta; a,b,c) &\equiv&    \frac{(x-a)^2+(y-b)^2+ \zeta^2 - c^2} { \sqrt{((x-a)^2+(y-b)^2+ \zeta ^2+ c^2)^2 -  4\, c^2\, \zeta^2 }} \nonumber\\ 
&=&    \frac{(x-a)^2+(y-b)^2+ \zeta^2 - c^2} { \sqrt{((x-a)^2+(y-b)^2+ (\zeta-c)^2) ((x-a)^2+(y-b)^2+ (\zeta+c)^2) }} \,. \nonumber\\
\label{Ddefn}
\end{eqnarray}
One can then take: 
\begin{equation}
V~=~ \varepsilon_0  ~+~ \sum_{j=1}^N \, q_j \,  G(x,y,\zeta; a_j,b_j,c_j)   
\label{Vblowup}
\end{equation}
\begin{equation}
A ~=~  \sum_{j=1}^N \, q_j \,  D(x,y,\zeta; a_j,b_j,c_j) \, d\phi   
\label{Ablowup}
\end{equation}
With these choices and $ \varepsilon_0=1$, the LeBrun-Burns metric is a smooth K\"ahler metric on $\IC^2$ blown up at $N$ points.  It is thus a K\"ahler, electrovac generalization of the Gibbons-Hawking metrics.
 
Near $(a_j,b_j,c_j)$, one has 
\begin{equation}
G(x,y,\zeta; a_j,b_j,c_j) ~\sim~     \frac{c_j} { \sqrt{ (x-a_j)^2+(y-b_j)^2+ (\zeta-c_j)^2 }} ~\equiv~ \frac{c_j}{r}\,,
\label{Gasympsing}
\end{equation}
\begin{equation}
D(x,y,\zeta; a_j,b_j,c_j) ~\sim~     \frac{\zeta - c_j} { \sqrt{ (x-a_j)^2+(y-b_j)^2+ (\zeta-c_j)^2 }} ~\equiv~ \cos \theta\,,
\label{Dasympsing}
\end{equation}
and the metric (\ref{LBB2}) behaves as:
\begin{eqnarray}
ds^2_4  &=&  c_j\, q_j \big[  q_j^{-2} r \, (d\tau+\cos \theta \, d\phi)^2 + r^{-1}   (dr^2 + r^2  d \theta^2+ r^2  \sin^2 \theta \, d \phi^2 ) \big]  \nonumber \\
 &=&  c_j\, q_j \big[ d\rho^2 +   \coeff{1}{4}  \rho^2(  d \theta^2+    \sin^2 \theta \, d \phi^2  +  q_j^{-2}  \, (d\tau+\cos \theta \, d\phi)^2) \big] \,, 
\label{metasymp}
\end{eqnarray}
where we have introduced spherical polar coordinates about $(a_j,b_j,c_j)$ and made a change of variable $r = {1\over 4} \rho^2$.  Thus near the singular points of $V$, the metric is locally $\IR^4/\ZZ_{q_j}$, and hence may be viewed as regular in string theory.

At infinity one has:
\begin{equation}
G(x,y,\zeta; a_j,b_j,c_j) ~\sim~     \frac{2\, c_j^2 \, \zeta^2 } {( x^2+ y^2+ \zeta^2)^2}\,,
\label{Gasympinf}
\end{equation}
\begin{equation}
D(x,y,\zeta; a_j,b_j,c_j) ~\sim~     1\,,
\label{Dasympinf}
\end{equation}
and hence $V \to  \varepsilon_0$ and $A \to d\phi$, and the metric is asymptotic to $\IR^4 = \IC^2$ for  $\varepsilon_0 = 1$.

\subsubsection{Homology and periods}
\label{Topology}

\begin{figure}
\centering
\includegraphics[height=3cm]{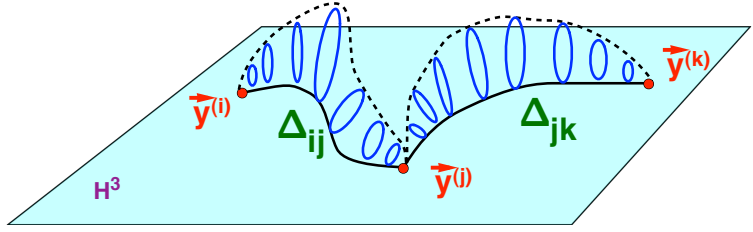}
    \caption{\it \small The non-trivial cycles of the LeBrun-Burns metrics are defined by sweeping the $U(1)$ fiber along a path, in $\IH^3$,  between 
 any two poles of the potential, $V$. The fiber is pinched off at the poles.  Here the fibers sweep out a pair of intersecting two-cycles.}
\label{cycles}
\end{figure}

Exactly as in Gibbons-Hawking geometries, the LeBrun-Burns metrics have non-trivial two-cycles defined by the $U(1)$ fibers over any curve between the poles of $V$.  More specifically,  the $U(1)$ fiber (defined by $\tau$) taken over a generic line interval in the $\IH^3$ base describes a cylinder.  However, if one runs this interval between two poles of $V$ at points, $\vec y^{(i)}$ and $\vec y^{(j)}$ then the fiber is pinched off at the ends and the result is essentially a topological two-sphere.   The asymptotic behavior of the metric at each end of the interval, (\ref{metasymp}),  means that this two-sphere may, in fact, be modded out by some discrete group that depends upon the values of $q_i$ and $q_j$.     The two-cycles defined in this way will be denoted as  $\Delta_{ij}$ and are depicted in Fig. \ref{cycles}.

The periods of these cycles are trivial to compute using (\ref{Kform}):
\begin{equation}
\frac{1}{2\, \pi} \, \int_{\Delta_{ij}} \, J ~=~ \frac{1}{2\, \pi} \,  \int_{\Delta_{ij}}  d\tau \wedge dz ~=~ 2(z_j  - z_i) ~=~ \zeta_j^2 - \zeta_i^2    \,,   \label{perint}
\end{equation}
where $z_i = \frac{1}{2}  \zeta_i^2$ denote the $z$-coordinates of the corresponding poles of $V$.

The Maxwell fields, $\Theta^{(1)}$, $\Theta^{(2)}$ and $\Theta^{(3)}$ defined in (\ref{Thetaforms}) and (\ref{Max3}) have components along the fiber of the form 
\begin{equation}
\Theta^{(I)} ~=~  \vec \nabla \bigg( \frac{K^{(I)}}{w} \bigg) \,\cdot \, d\tau \wedge d\vec y \,,   \qquad I=1,2,3 \,, \label{tauparts}
\end{equation}
where $K^{(1)}$ and $K^{(2)}$ satisfy (\ref{lin2a}) and   (\ref{lin2b}) and 
\begin{equation}
K^{(3)} ~\equiv~    {1 \over 2} \,  \partial_z u  \,.  \label{p3defn}
\end{equation}
From this it follows that these fields have fluxes 
\begin{equation}
\Pi_{ij}^{(I)}  ~\equiv~   \frac{K^{(I)}}{w}\Big|_{\vec y^{(j)}} ~-~ \frac{K^{(I)}}{w} \Big |_{\vec y^{(i)}} \,, \qquad  I=1,2,3 \,.  \label{fluxes}
\end{equation}
Note, in particular, that for the LeBrun-Burns metric ${K^{(3)}}w^{-1} =   V^{-1}$ which vanishes at  all the $\vec y^{(i)}$.  Therefore $\Theta^{(3)}$ has {\it no non-trivial} fluxes on the compact two cycles.  

On the other hand, the complete Maxwell field, $\cF$, (\ref{MaxF}),  does have non-trivial fluxes because it has an anti-self-dual component involving $J$.  
\begin{equation}
\frac{1}{2\, \pi} \, \int_{\Delta_{ij}} \, \cF ~=~ -\frac{1}{2\, \pi} \, \int_{\Delta_{ij}} \, J   ~=~ -(\zeta_j^2 - \zeta_i^2)    \,,   \label{Fflux}
\end{equation}
There is potentially a similar contribution from the ``electric parts''  of the complete gauge fields defined in (\ref{AAnsatz}).  Specifically, the complete Maxwell fields, $F^{(I)}$, have a component along the fiber:
\begin{equation}
 F^{(I)}  ~\equiv~ dA^{(I)} ~=~   d\tau \wedge d \big(Z_I^{-1} \, \mu \big) ~+~ \dots \,.  \label{Ftaucomps}
\end{equation}
This will generically lead to fluxes of the form
\begin{equation}
\widehat \Pi_{ij}^{(I)}  ~\equiv~   \frac{\mu}{Z_I}\bigg |_{\vec y^{(j)}} ~-~ \frac{\mu}{Z_I} \bigg |_{\vec y^{(i)}} \,, \qquad  I=1,2,3 \,,  \label{otherfluxes}
\end{equation}
however, regularity of the metric and the absence of closed time-like curves typically requires that $Z_I$ be finite and $\mu$ vanish at the points $\vec y^{(i)}$. Thus these fluxes are usually zero.  

In summary, the bubbled non-BPS solutions generically have non-vanishing fluxes for all three Maxwell fields but the self-dual part of the third Maxwell field, $\Theta^{(3)}$, will have trivial fluxes.

\subsection{Solving the non-BPS system}
\label{LBBSoln}

The differential operators of interest are:
\begin{eqnarray}
\cL_1 H & \equiv&  \partial_{x}^2 H   ~+~ \partial_{y}^2 \, H  ~+~ \zeta^{-1} \, \partial_\zeta \, (\zeta  \partial_\zeta  H )   \,,  \label{diffop1} \\
\cL_2 G & \equiv&  \partial_{x}^2 G   ~+~ \partial_{y}^2 \, G  ~+~\zeta \, \partial_\zeta \, \big (\zeta^{-1} \,  \partial_\zeta  G \big) \,. \label{diffop2}
\end{eqnarray}
Note that $\zeta^2 \cL_2$ is simply the Laplacian on $\IH^3$.  The operator, $\cL_1$ has been introduced for later convenience because it appears in the equations  of motion and it is also useful to note that it has a simple geometric interpretation.   Observe that the Laplacian on $\IR^4 = \IR^2 \times \IR^2$ may be written as  
\begin{equation}
\hat \cL_1 H  ~=~   \partial_{x}^2 H   ~+~ \partial_{y}^2 \, H  ~+~ \zeta^{-1} \, \partial_\zeta \, (\zeta  \partial_\zeta  H )   ~+~ \zeta^{-2}\, \partial_\varphi^2 \, H\,,  \label{diffop3}
\end{equation}
where $(x,y)$ are Cartesians on the first $\IR^2$ and $(\zeta,\varphi)$ are polars on the second $\IR^2$.  Thus solving equations that involve $\cL_1$ may simply be viewed as looking for $\varphi$-independent solutions with the flat Laplacian on $\IR^4$.  The equations and solutions that involve $\cL_1$ are thus extremely familiar from the extensive literature on black rings. In particular, it is useful to note that the following are Green functions for $\cL_1$:
\begin{eqnarray}
H(x,y,\zeta; a,b,c) &\equiv&    \frac{1} { \sqrt{((x-a)^2+(y-b)^2+ \zeta ^2 + c^2)^2 -  4\, c^2\, \zeta^2 }}  \nonumber\\
&=&  \frac{1} { \sqrt{((x-a)^2+(y-b)^2+ (\zeta-c)^2) \, ((x-a)^2+(y-b)^2+ (\zeta+c)^2) }} \,. \nonumber\\
\label{Fdefn}
\end{eqnarray}
At infinity one has:
\begin{equation}
H(x,y,\zeta; a,b,c) ~\sim~     \frac{1} {( x^2+ y^2+ \zeta^2) }\,.
\label{Fasympinf}
\end{equation}

To solve the linear system, one first solves the homogeneous equations:
\begin{equation}
\cL_2 V   ~=~  0  \,,  \qquad  \cL_2 \, (\zeta^2 \, L_1)    ~=~   0   \,,  \qquad  \cL_2  (\zeta^2 \, L_2)   ~=~ 0   \,, \label{VLeqns}  
\end{equation}
and then uses these solutions in the equations that define the magnetic fluxes and the angular momentum function, $M$:
\begin{eqnarray}
  \cL_1 \, K^{(1)}    &=&   2\, \zeta^{-1} \partial_\zeta  \,(V\,   L_2 )   \,,  \qquad  \cL_1 \, K^{(2)} ~=~  2\, \zeta^{-1} \partial_\zeta \,(V \,   L_1 ) \,, \label{peqns}   \\
 \cL_1 \, M  &=&  \zeta^{-1} \partial_\zeta \, (\zeta^2 \,  L_1 \, L_2 ) \,.  \label{Meqn1}   
\end{eqnarray}
The last step is to use these solutions in:
\begin{equation}
\cL_2 \, L_3  ~=~    4 \,    V \,( L_1 L_2 -  \zeta^{-1} \partial_\zeta  M) -  2 \,  \zeta \, (L_1 \, \partial_\zeta K^{(1)} + L_2 \,\partial_\zeta  K^{(2)} )\,. \label{L3eqn1}
\end{equation}

The physical functions now have the form
\begin{equation}
Z_1 ~=~  {K^{(2)}  \over V}  ~+~L_1\,, \quad Z_2 ~=~ { K^{(1)} \over V }  ~+~L_2 \,, \quad Z_3 ~=~    {\zeta^2 \,K^{(1)} \,  K^{(2)}  \over  V }  ~+~L_3   \,, \label{BurnsZforms}
\end{equation}
\begin{equation}
\mu ~=~   -{\zeta^2 \,  K^{(1)} \,  K^{(2)}  \over V^2}  - \frac{1}{2}\,   {\zeta^2 \, (K^{(1)} \,  L_1  + K^{(2)} \,  L_2 )  \over V}  ~-~ \frac{1}{2}\,    { L_3  \over V}  ~+~ M \,.  \label{Burnsmuform}
\end{equation}
The equations for $\omega$ reduce to
\begin{eqnarray}
&& (\partial_y \omega_{\zeta} -  \partial_{\zeta}  \omega_y ) + \ds\frac{1}{\zeta}(M \partial_x V - V \partial_x M) + \frac{1}{2\zeta} \sum_{j=1}^2 (K^{(j)} \partial_x (\zeta^2 L_j) -  \zeta^2L_j  \partial_x K^{(j)}) + \frac{1}{2 \zeta}  \partial_x L_3  = 0 \,, \notag \\
&& (\partial_{\zeta}  \omega_x -\partial_x \omega_{\zeta}) + \ds\frac{1}{\zeta}(M \partial_y V - V \partial_yM) + \frac{1}{2\zeta} \sum_{j=1}^2 (K^{(j)} \partial_y (\zeta^2 L_j) -  \zeta^2L_j  \partial_y K^{(j)}) + \frac{1}{2 \zeta}  \partial_y L_3  = 0 \,, \notag \\
&& (\partial_{x}  \omega_y -\partial_y \omega_{x}) + \ds\frac{1}{\zeta}(M \partial_{\zeta} V - V \partial_{\zeta} M) + \frac{1}{2\zeta} \sum_{j=1}^2 (K^{(j)} \partial_{\zeta} (\zeta^2 L_j) -  \zeta^2L_j  \partial_{\zeta} K^{(j)}) \label{Burnsomega}\\ 
&&+ \frac{1}{2 \zeta}  \partial_{\zeta} L_3 + 2 V L_1L_2 = 0 \,. \notag 
 \end{eqnarray}
This system of equations has a gauge invariance that leaves the physical solution completely invariant.  See Appendix \ref{gauge invariance} for details.

\subsection{Asymptotics}
\label{Asymp1}

Ideally one  would like to find solutions that are asymptotically flat and for the metric (\ref{metAnsatz}) this means that one must have $Z_I \to 1$ at infinity.  However, this is generically not possible because the Maxwell fields, $\cF$, in (\ref{Fdecomp})  and $\Theta^{(j)}, j=1,2$  in (\ref{Thetaforms})   
 involve the K\"ahler form and thus the norm of these fields does not vanish at infinity.  
 
To see this, suppose that the $Z_I$ go to constants at infinity, then the space-space and time-time components of the full five-dimensional Einstein's equations imply that, at infinity, 
\begin{equation} \label{5d_Einstein_ij}
\sum_{I=1}^3 \big( {\Theta^{(I)}}_{ik} {\Theta^{(I)}}_j{}^k - \frac14 \delta_{ij} {\Theta^{(I)}}_{kl} {\Theta^{(I)}}^{kl} \big) ~\to~ 0 \,,
\end{equation}
\begin{equation} \label{5d_Einstein_00}
\sum_{I=1}^3 {\Theta^{(I)}}_{kl} {\Theta^{(I)}}^{kl} ~\to~  0 \,.
\end{equation}
The self-dual parts of the $\Theta^{(I)}$ can be made to vanish at infinity, and so \eqref{5d_Einstein_ij} can be satisfied.  However, the left-hand-side of \eqref{5d_Einstein_00} is positive-definite, and so this equation cannot be satisfied because the $\Theta^{(j)}, j=1,2$ always have a non-vanishing anti-self-dual part (given by the K\"ahler form $J$).   This is in contrast to the solutions based on Gibbons-Hawking metrics for which the $\Theta^{(I)}$ vanish at infinity.  

One cannot, therefore, arrange to have $Z_I \to 1$ at infinity.   On the other hand, we will now show that there are solutions that are asymptotic to the near-horizon limit of a BMPV black hole.

\section{Solutions with a flat base space}
\label{FlatSolutions}

Here we consider solutions in which the base space is completely flat, taking $V \equiv 1$ in the LeBrun-Burns metric.  One of the purposes in doing this is to see 
what kind of asymptotic geometries and black-object geometries can be generated from the  LeBrun-Burns metric using the solution technique of Section \ref{BurnsSec}. 
Indeed, we will show that the natural boundary conditions correspond to the near-horizon regions of black holes and black rings.   It is also important to note that even though we have set $V=1$ and thus trivialized the metric on the base,  the Maxwell field, $\cF$, is still non-zero but is now purely anti-self-dual\footnote{This means that $\cF$ has vanishing energy-momentum tensor, consistent with the flatness of the base.}  and proportional to the complex structure, $J$.  Similarly, the other Maxwell fields (\ref{Thetaforms}) have both anti-self-dual and self-dual parts on the base.  This will generically mean that supersymmetry is completely broken and that the solutions we get will be non-BPS.   

\subsection{The near-horizon limit of a black hole}
\label{flatsol1}

Perhaps the simplest non-trivial solution is a spherically symmetric one, whose sources necessarily lie at $(x, y, \zeta)=(0,0,0)$.  In addition we set some of the electric potentials to zero: 
\begin{equation}
L_1 ~\equiv~L_2 ~\equiv~ 0\,. \label{BHsol1}
\end{equation}
It is also convenient to introduce polar coordinates in $\IR^2$ and $\IR^4$:  We already have $\zeta$ and $\tau$ in one copy of $\IR^2$ and so we define\footnote{The coordinate $\theta$ here is not the same as the one in \eqref{metasymp}. }
\begin{equation}
x ~=~ \eta \cos \phi\,, \quad y ~=~ \eta \sin \phi\,;   \qquad      \zeta ~=~ \rho \cos \theta \,, \quad  \eta ~=~ \rho \sin \theta  \,; \qquad \rho ~\equiv~  x^2+y^2+ \zeta^2 \,.  \label{polars1}
\end{equation}

The functions $K^{(I)}$ and $M$ are then homogeneous solutions to  $\cL_1 H =0$ and the spherically symmetric solutions are proportional to $H(x,y,\zeta; 0,0,0) = \rho^{-2}$    (see  (\ref{Fdefn})).  We therefore take
\begin{equation}
  Z_1 ~=~  K^{(2)} ~=~   \frac{\beta_2}{  \rho^2}  \,, \qquad Z_2 ~=~   K^{(1)} ~=~      \frac{   \beta_1}{  \rho^2} \,,  \qquad M ~=~   \frac{\gamma}{  \rho^2}  \label{BHsol2}  \,, 
\end{equation}
where  $\beta_1, \beta_2 $ and $\gamma$ are constant parameters.  

It is easy to see that one can satisfy (\ref{L3eqn1}) by taking:
\begin{equation}
L_3  ~=~\hat L_3 + 2 \, M \,, \qquad \cL_2 \hat L_3  ~=~  0\,, \label{BHsol3}
\end{equation}
for some function, $\hat L_3$.    The natural choice for $\hat L_3$ is the function $G$ in (\ref{Gdefn}), but this vanishes for $c=0$, and one must take a limit:
\begin{equation}
 \hat L_3  =   \beta_3 \, \lim_{c \to 0} \, \frac{ 1}{ 2\,c^2} \,  G(x,y,\zeta; 0,0,c) ~=~  \beta_3 \, \frac{ \zeta^2}{ \rho^4}  ~=~  \beta_3 \, \frac{ \cos^2 \theta}{ \rho^2}\,,  \label{BHsol4}
\end{equation}
One then has
\begin{eqnarray}
 Z_3 &=&  \zeta^2 \, K^{(1)} \,   K^{(2)} + L_3 ~=~  (\beta_1 \,\beta_2\,  +  \beta_3 )\,  \frac{ \cos^2 \theta}{ \rho^2}  ~+~      \frac{2\, \gamma}{  \rho^2} \,,  \label{BHsol5} \\
 \mu &=&   -\,  \zeta^2 \,  K^{(1)} \,  K^{(2)}     ~-~ \frac{1}{2}\, \hat  L_3 ~=~ - \frac{1}{2}\, (2\,\beta_1 \,\beta_2 +   \beta_3) \,   \frac{ \cos^2 \theta}{ \rho^2} \,.  \label{BHsol6}
\end{eqnarray}

The last step is to solve for $\vec \omega$, for which we can choose the gauge $\omega_z =0$.  Equations (\ref{omx})--(\ref{omz}) then reduce to:
\begin{equation}
\zeta \,  \partial_\zeta \omega_y  ~=~    \frac{1}{2}\, \partial_x  \hat L_3   \,,  \quad      \zeta \,  \partial_\zeta \omega_x  ~=~   - \frac{1}{2} \, \partial_y \hat L_3   \,,  \qquad  \partial_x \omega_y - \partial_y \omega_x ~=~ - \frac{1}{2}\, \zeta^{-1} \,  \partial_\zeta \hat L_3  \label{BHsol7}
\end{equation}
for which the solution is:
\begin{equation}
\omega ~=~   - \frac{\beta_3}{2}  \, \frac{1}{\rho^4} \,( y \, dx -x \, dy )~=~      \frac{\beta_3}{2}   \, \frac{ \sin^2 \theta}{ \rho^2} \, d \phi\,,  \label{BHsol8}
\end{equation}
where the homogeneous solutions have been chosen so that $\omega$ goes to zero at infinity.

The five-dimensional metric is then:
\begin{eqnarray}
ds_5^2  &=& - W_0(\theta)^{-2} \, \rho^4\,\bigg(dt - \frac{1}{2}\, (  \beta_3+  2\,\beta_1 \,\beta_2 ) \,   \frac{ \cos^2 \theta}{ \rho^2} d\tau    +  \frac{\beta_3}{2}   \,\frac{ \sin^2 \theta}{ \rho^2} \, d \phi \bigg)^2 \nonumber \\
&&\quad ~+~ W_0(\theta) \, \Big(  \frac{d\rho^2}{ \rho^2}  + d \theta^2 + \cos^2 \theta d\tau^2  +  \sin^2 \theta  d \phi^2  \Big)  \,,
\label{BHmet}
\end{eqnarray}
where 
\begin{equation}
W_0(\theta)  ~\equiv~   \big(\beta_1 \beta_2(  2\, \gamma+ (\beta_1 \,\beta_2\,  +  \beta_3 )\, \cos^2 \theta  ) \big)^\frac{1}{3}  \,.  \label{BHsol9}
\end{equation}
The conditions for absence of causal pathologies for solutions of our Ansatz are discussed in Appendix \ref{causality}. For the simple solution in this section there is no Dirac-Misner string in $\omega$ and the condition for absence of CTC's is that all constants $\gamma$, $\beta_1$, $\beta_2$ are non-negative and
 \begin{equation}
 8 \gamma\beta_1\beta_2 \geq \beta_3~. \label{noCTCBMPV}
 \end{equation}
For generic choice of parameters satisfying \eqref{noCTCBMPV} the metric \eqref{BHmet} has the form of a warped rotating $AdS_2\times S^3$. The general solution has unequal angular momenta in each $\IR^2$, and has a distorting warp factor, function, $W_0(\theta)$.  For the special choice $\beta_3 = - \beta_1 \beta_2$ the function $W_0$  becomes a constant and the two angular momenta become equal. The metric then is precisely the near horizon limit of the BMPV black hole \cite{Breckenridge:1996is}. It is worth emphasizing that the BMPV black hole (and its near horizon limit) is a supersymmetric solution of supergravity whereas our solution has anti-self-dual flux that breaks supersymmetry.

\subsection{Near-horizon  limit of a black ring with two dipole charges}
\label{flatsol2}

\subsubsection{Solving the equations}
\label{SolEqns}

To get the black ring (or supertube) generalization of the foregoing solution, we simply need to use the source functions (\ref{Gdefn}) and  (\ref{Fdefn})  with $c \ne 0$.    We can, without loss of generality, take $a=b=0$ since we are going to consider a single source.   We therefore start by taking: 
\begin{equation}
 \qquad  L_J ~=~  \ds\frac{\ell_J}{\zeta^2} \,  G(x,y,\zeta; 0,0,c) \,,  \qquad  J=1,2 \,, \label{BRsol1}
\end{equation}
for some constants $\ell_J $.   The supertube will thus be located at $(0,0,c)$.

The functions $K^{(I)}$ now have a source part and a homogeneous  part that is proportional  to  $H$:
\begin{eqnarray}
K^{(1)} &=&   - \frac{\ell_2}{\zeta^2}\,   G(x,y,\zeta; 0,0,c)   +  \beta_1 \, H(x,y,\zeta; 0,0,c)  ~=~  - L_2  +  \beta_1 \, H(x,y,\zeta; 0,0,c)   \,,  \label{BRsol2}  \\
K^{(2)} &=&  - \frac{\ell_1}{\zeta^2}\,   G(x,y,\zeta; 0,0,c)   + \beta_2 \, H(x,y,\zeta; 0,0,c)  ~=~  - L_1  + \beta_2 \, H(x,y,\zeta; 0,0,c)  \label{BRsol3}  \,, 
\end{eqnarray}
and hence
\begin{equation}
 Z_1 = K^{(2)} + L_1 =   \beta_2\,  H(x,y,\zeta; 0,0,c)   \,, \qquad Z_2 =   K^{(1)} + L_2  =    \beta_1\,  H(x,y,\zeta; 0,0,c)  \,, \label{BRsol4} \\
\end{equation}
where, once again, $\beta_1$ and $\beta_2 $ are constants.  The next step is to solve (\ref{Meqn1})  and   (\ref{L3eqn1})   for $M$ and $L_3$, and as in  (\ref{BHsol3}) it is convenient to shift $L_3$ by $M$.  One finds
\begin{equation}
L_3  ~=~\hat L_3 + 2 \, M \,, \qquad \cL_2 \hat L_3  ~=~ -2\, (\beta_1\ell_1 + \beta_2 \ell_2) \,  \zeta^{-1} \, G \, \partial_\zeta H  \,. \label{BRsol5}
\end{equation}
It is straightforward to show that this and  (\ref{Meqn1})  are satisfied by:
\begin{eqnarray}
\hat L_3  &=&   (\beta_1\ell_1 + \beta_2 \ell_2) \, \big[\, (\rho^2 + c^2 - 2 \, \zeta^2) H^2 ~-~H\,\big]    +  \ell_3   \, G   \,,   \label{BRsol6}  \\
M  &=&   - \frac{ \ell_1 \, \ell_2}{2}   \, \big( \zeta^{-2} \, G^2  ~-~   4 \, \rho^2\, H^2  \label{BRsol7} \big) ~+~ \mu_0\,  H  \,, 
\end{eqnarray}
where the constants $\ell_3$ and $\mu_0$ multiply homogeneous solutions to the relevant differential equations.  Combining these results in 
 (\ref{Z3form}) and  (\ref{muform})   one obtains: 
\begin{eqnarray}
 Z_3 &=&   \big((2\,\ell_2-\beta_1) (2\,\ell_1-\beta_2) \zeta^2   ~+~ 4\, \ell_1\,\ell_2 (\rho^2 -\zeta^2)  \big)\, H^2 + \ell_3 \, G  ~+~ 2\, \mu_0 \, H\,,      \label{BRsol8} \\
 \mu &=&   - \frac{1}{ 2}\, \big[2\, (\beta_1 \beta_2 - (\beta_1\ell_1 + \beta_2 \ell_2) )  \,    \zeta^2 \, H^2     ~+~  \ell_3 \, G \big]  \,.  \label{BRsol9}
\end{eqnarray}

The last step is to solve for (\ref{omx})--(\ref{omz}) for $\omega$ and one can easily verify that: 
\begin{equation}
\omega =  \Big[ - \frac{\ell_3}{2} \big(  G - 2\, c^2 \, H \big)  ~-~  (\beta_1\ell_1 + \beta_2 \ell_2)  \,  H^2 \, \rho^2 \sin^2 \theta \Big] \,  d \phi   \,,  \label{BRsol10}
\end{equation}
where a constant of integration has been adjusted so that $\omega$ goes to zero at infinity.  

The Green functions, $G$ and $H$, and the metric that we construct here are, of course, familiar from the standard description of black rings.  This is explained further in Appendix \ref{CanCoords}.

\subsubsection{Regularity near the supertube}
\label{STreg}

The first step in checking regularity of the metric is to look for closed time-like curves (CTC's)  in the $\tau$ and $\phi$ directions near the  ring.  Indeed there is a divergent negative coefficient of $d \tau^2$ unless one requires:
\begin{equation}
 \beta_1 \, \ell_1   ~=~  \beta_2 \, \ell_2\,, \label{CTCcond1}
\end{equation}
which we will use to eliminate $\ell_2$.  This condition is a familiar regularity requirement for the three-charge, two-dipole charge supertube 
\cite{Bena:2004wt,Bena:2004wv}.

Once this term is dealt with there is a sub-leading divergence that then requires 
\begin{equation}
\beta_2 \mu_0 =  - 2 c^2 \ell_1 \ell_3~. \label{radiusST}
\end{equation}
Since the radius of the supertube\footnote{See Appendix A.4 for more details on this.} is related to $c$ we can interpret this relation as a radius formula for the supertube. There is a Dirac-Misner string in $\omega$ that leads to CTC's for $x = y =0$ and $\zeta < c$, and this requires that $\ell_3 =0$ and  so we must fix the homogeneous solutions by taking:
\begin{equation}
 \ell_3   ~=~  \mu_0 ~=~ 0   \,. \label{CTCcond2}
\end{equation}
With these choices the five-dimensional metric simplifies significantly and it is convenient to introduce two manifestly non-negative functions:
\begin{equation}
W_1(\rho, \theta)    \equiv   \rho^4 \,\big( (\rho^2+ c^2)^2  - 4 \, c^2 \, \rho^2\,\cos^2  \theta \big)^{-1}   \,, \quad   W_2(\theta)   \equiv  (2\, \ell_1-\beta_2)^2 \, \cos^2 \theta  + 4 \, \ell_1^2 \, \sin^2  \theta \,,  \label{H12defn}
\end{equation}
and define
\begin{equation}
\widehat Z(\rho,\theta)    ~\equiv~ \bigg(\beta_1^2 \, W_1^2 \, W_2\bigg)^{1/3}  \,. \label{Zhatdefn}
\end{equation}
The angular momentum vector simplifies to:
\begin{equation}
k   ~=~  \rho^{-2}\,   W_1(\rho, \theta)    \, \Big[ \big( (\beta_1\ell_1 + \beta_2 \ell_2) - \beta_1\, \beta_2\,    \big) \, \cos^2 \theta \, d \tau ~-~ (\beta_1\ell_1 + \beta_2 \ell_2)  \, \sin^2 \theta \, d \phi  \Big]   \,, \label{kang1}
\end{equation}
and the five dimensional metric may be written as: 
\begin{equation}
ds_5^2  ~=~  - \widehat Z(\rho,\theta)^{-2} \, \rho^4\, (dt + k)^2 ~+~  \widehat Z(\rho,\theta)   \, \Big(  \frac{d\rho^2}{ \rho^2}  + d \theta^2 + \cos^2 \theta \, d\tau^2  +  \sin^2 \theta  \, d \phi^2  \Big)  \,.
\label{BRmet}
\end{equation}
This form of the metric is reminiscent of the ``decoupling limit" for supersymmetric black rings/supertubes discussed in \cite{Elvang:2004ds}. It would be interesting to dualize our solutions to the D1-D5 duality frame and compare in more detail with the background  in \cite{Elvang:2004ds}.

This metric has no CTC's but is singular at the location of the supertube.  This is a generic property of the supertube with two dipole charges and is familiar from the corresponding supersymmetric solutions \cite{Bena:2004wt}.

\subsubsection{Asymptotics at infinity}
\label{AsympInf}

At infinity one has:
\begin{equation}
Z_1 ~\sim~    \frac{\beta_2}{\rho^2}  \,, \qquad Z_2~\sim~    \frac{\beta_1}{\rho^2}    \,, \qquad Z_3~\sim~      \frac{\beta_1}{\beta_2} \, \frac{W_2(\theta)}{\rho^2}      \,,  \label{asymp1}
\end{equation}
\begin{equation}
k   ~\sim~  \rho^{-2}\,  \Big[  \big( (\beta_1\ell_1 + \beta_2 \ell_2) - \beta_1\, \beta_2\,    \big) \, \cos^2 \theta \, d \tau ~-~ (\beta_1\ell_1 + \beta_2 \ell_2)  \, \sin^2 \theta \, d \phi  \Big]   \,, \label{asymp2}
\end{equation}
and hence:
\begin{equation}
\big( Z_1 Z_2 Z_3 \big)^{1 \over 3}  ~\sim~ \,   \frac{\Big( \beta_1^2 \, W_2(\theta) \Big)^{1 \over 3}}{\rho^2}   \label{Warpasymp1}
\end{equation}
The five-dimensional asymptotic metric, (\ref{BRmet}),  now takes the form:
\begin{eqnarray}
 ds_5^2  ~=~  &&  - \Big(\beta_1^2 \, W_2(\theta) \Big)^{-2/3} \, \rho^4\, (dt + k)^2 \nonumber \\
&&\quad ~+~ \Big(\beta_1^2  \, W_2(\theta) \Big)^{1/3}\, \Big(  \frac{d\rho^2}{ \rho^2}  + d \theta^2 + \cos^2 \theta \, d\tau^2  +  \sin^2 \theta  \, d \phi^2  \Big)  \,,
\label{BRmetasymp}
\end{eqnarray}
which is precisely of the form discussed in Section \ref{flatsol1},  {\it i.e.} a warped form of rotating $AdS_2 \times S^3$. For the special choice $\beta_2=4\ell_1$ we have $W_2(\theta)=4\ell_1^2$ and \eqref{BRmetasymp} reduces to the near horizon BMPV metric. Therefore we can consider the metric \eqref{BHmet} as the metric to which all black hole, black ring and multi-center solutions within our Ansatz will asymptote for $\rho\to\infty$. We will describe this in more detail in the next section.

\section{Multi-centered solutions}
\label{BlowingBubbles}

We have been able to solve in complete generality the system of differential equation \eqref{VLeqns}, \eqref{peqns}, \eqref{Meqn1}, \eqref{L3eqn1} and\eqref{Burnsomega} on an axisymmetric LeBrun-Burns base. This provides an infinite class of explicit five-dimensional multi-centered solutions with (at least) one time-like and two space-like Killing vectors $(\partial_{t},\partial_{\tau},\partial_{\phi})$. Amongst our solutions are multi-center generalizations of the solutions in Section \ref{flatsol1} and  \ref{flatsol2}  as well as a class of regular bubbled geometries that we discuss in some detail in Section \ref{Burnsasymptotics} below. 

\subsection{General axisymmetric solutions}
\label{GenAxi}

We will look for solutions on an axisymmetric LeBrun-Burns base in which the geometry at infinity has the form (\ref{BHmet}). The singular points of the harmonic function, $V$, that determines the LeBrun-Burns base are located along the $\zeta$ axis at points $c_j$:
\begin{equation}
V~=~ \varepsilon_0  ~+~ \sum_{j=1}^N \, q_j \,  G_j ~. 
\end{equation}
Where for convenience we have defined
\begin{eqnarray}
G_{i} &\equiv& G(x,y,\zeta;0,0,c_i)   = \ds\frac{\rho^2 +c_i^2}{\sqrt{(\rho^2+c_i^2)^2-4 \zeta^2 c_i^2}} ~-~ 1, \\
H_{i} &\equiv& H(x,y,\zeta;0,0,c_i)   = \ds\frac{1}{\sqrt{(\rho^2+c_i^2)^2-4 \zeta^2 c_i^2}}~,\\
D_{i} &\equiv& D(x,y,\zeta; 0,0,c_i) =  \ds\frac{\rho^2 -c_i^2}{\sqrt{(\rho^2+c_i^2)^2-4 \zeta^2 c_i^2}}~, 
\end{eqnarray}
where we will assume that $c_i\neq 0$. It is trivial to solve \eqref{VLeqns} for the functions $L_1$ and $L_2$
\begin{equation}
L_{a} = \ds\frac{1}{\zeta^2} \left(\ell_a^{0} +\ds\sum_{i=1}^{N}\ell_{a}^{i}G_{i}\right)~, \qquad a=1,2~. \label{L12axi}
\end{equation}
Solving \eqref{peqns} and \eqref{Meqn1} for $K^{(a)}$ and $M$ one finds
\begin{eqnarray}
K^{(1)} &=& k_1^{0} + \ds\frac{\beta_1}{\rho^2}+\ds\sum_{i=1}^{N}k_{1}^{i}H_{i} - V L_{2}+ 4\rho^2 \ds\sum_{i,j=1}^{N} q_{i} \ell_2^{j}H_iH_j~, \label{K1axi}\\
K^{(2)} &=& k_2^{0} + \ds\frac{\beta_2}{\rho^2}+\ds\sum_{i=1}^{N}k_{2}^{i}H_{i} -  V L_{1} + 4\rho^2 \ds\sum_{i,j=1}^{N} q_{i} \ell_1^{j}H_iH_j~, \label{K2axi}\\
M &=& m_{0}+ \ds\frac{\gamma}{\rho^2} +\ds\sum_{i=1}^{N}m_{i}H_{i} -  \ds\frac{\zeta^2}{2} L_{1}L_2 + 2\rho^2 \ds\sum_{i,j=1}^{N} \ell_1^{i} \ell_2^{j}H_iH_j~.\label{Maxi}
\end{eqnarray}
After a somewhat tedious exercise\footnote{Some of the identities used to solve the equations for $K^{(a)}$, $M$, $L_3$ and $\omega_{\phi}$ are collected in Appendix \ref{UseIdents}.} one can also solve equation \eqref{L3eqn1} 
\begin{eqnarray} 
\label{L3axi}
&& L_3 = \ell_3^{0} + \ds\sum_{i=1}^{N} \ell_3^i G_i - \zeta^2 VL_1 L_2 + \ds\sum_{i=1}^{N} (2(\varepsilon_0 - Q)m_{i} +(\ell_1^0-\Lambda_1) k_1^i+(\ell_2^0-\Lambda_2) k_2^i)H_{i} \notag\\
&&+ \beta_3\ds\frac{\zeta^2}{\rho^4}+(2(\varepsilon_0 - Q)\gamma +(\ell_1^0-\Lambda_1) \beta_1+(\ell_2^0-\Lambda_2)\beta_2)\ds\frac{1}{\rho^2}+ 2\gamma \ds\sum_{i=1}^{N} \ds\frac{q_{i} }{c_i^2} \ds\frac{\rho^{-2}-H_i}{H_i}~, \notag\\
&&+ \ds\sum_{i=1}^{N} (2q_{i}m_{i} + \ell_1^i k_1^i + \ell_2^i k_2^i) (\eta^2-\zeta^2+c_i^2)H_{i}^2 + \ds\sum_{i\neq j =1}^{N} \ds\frac{(2q_{i}m_{j} + \ell_1^i k_1^j + \ell_2^i k_2^j)}{c_i^2-c_j^2} \ds\frac{H_j-H_i}{H_i} \notag\\
&&+4\ds\sum_{i,j=1}^{N} ((\varepsilon_0 - Q) \ell_1^i \ell_2^j+ (\ell_1^0-\Lambda_1) q_i \ell_2^j + (\ell_2^0-\Lambda_2)q_i \ell_1^j)\rho^2 H_i H_j \\ 
&&+ 4 \ds\sum_{i,j,k=1}^{N} q_i \ell_1^j \ell_2^k \rho^2 (3\rho^2 - 4 \zeta^2+c_i^2+c_j^2+c_k^2) H_{i} H_j H_k~. \notag
\end{eqnarray}
where we have defined
\begin{equation}
Q \equiv \ds\sum_{i=1}^{N}q_i~, \qquad\qquad \Lambda_{1} \equiv \ds\sum_{i=1}^{N} l_1^{i}~,  \qquad\qquad \Lambda_{2} \equiv \ds\sum_{i=1}^{N} l_2^{i}~.
\end{equation}
The one form $\omega = \omega_{\phi} d\phi$ is given by
\begin{eqnarray} 
\omega_{\phi} &=& \omega_{0}+ \ds\frac{\beta_3}{2} \ds\frac{\sin^2\theta}{\rho^2} - \gamma\ds\sum_{i=1}^{N} \ds\frac{q_i}{c_i^2} D_i - \ds\sum_{j=1}^{N} \left( m_0q_j + k_1^0 \ell_1^j + k_2^0 \ell_2^j + \ds\frac{\ell_3^j}{2} \right)D_j\notag\\ &&-  \ds\sum_{j=1}^{N} (2m_jq_j+k_1^j \ell_1^j + k_2^j \ell_2^j)\eta^2 H_j^2   - \ds\sum_{i\neq j =1}^{N} \ds\frac{(2q_{i} m_{j}+k_1^i \ell_1^j + k_2^i \ell_2^j)}{2(c_i^2-c_j^2)} (D_iD_j + 4\eta^2c_i^2 H_i H_j)  \notag \\
&& - 8 \ds\sum_{i,j,k=1}^{N} q_i \ell_1^j \ell_2^k \eta^2\rho^2 H_iH_jH_k  ~, \label{omegasoln}
\end{eqnarray}
where $\omega_{0}$ is a constant which should be fixed so as to avoid CTCs and Dirac-Misner strings.

Substituting \eqref{L12axi}, \eqref{K1axi}, \eqref{K2axi}, \eqref{Maxi} and \eqref{L3axi} in the expressions for $Z_1$, $Z_2$, $Z_3$ and $\mu$, \eqref{BurnsZforms} and \eqref{Burnsmuform}, one finds the most general non-BPS solution on an axisymmetric LeBrun-Burns base captured by the floating brane Ansatz of \cite{Bena:2009fi}. For easy comparison with the solution in Section \ref{flatsol1} we have chosen to single out the terms in the solution which have poles at $\rho=0$, {\it i.e.} the terms with coefficients involving $\beta_1$, $\beta_2$, $\beta_2$ and $\gamma$.

In addition to the parameters $\beta_1$, $\beta_2$, $\beta_3$ and $\gamma$, the solution in general has $(8N+7)$ parameters:  $\{c_i,\varepsilon_0, q_i, \ell_I^0, \ell_I^i, k_a^0, k_a^i,m_0, m_i\}$. As we will see in the next subsection imposing regularity and absence of causal pathologies will greatly reduce the number of independent parameters.

\subsection{Regular bubbled solutions}
\label{Burnsasymptotics}

The solution we construct here will be asymptotic to the metric \eqref{BHmet}, which can be viewed as the ``elementary" solution within our Ansatz. These regular solutions on a base with non-trivial topology can be viewed as a non-supersymmetric generalization of the BPS bubbled solutions of \cite{Bena:2005va, Berglund:2005vb}.

We begin by defining a radial coordinate around each of the poles of the harmonic functions
\begin{equation}
\rho_i^2= \eta^2 + (\zeta-c_i)^2~. 
\end{equation}
We will be interested in constructing a solution that is regular at the locations of the poles of the harmonic functions, $\rho_i \to 0$, and is free of CTCs and Dirac-Misner strings.

For $\rho_i \to 0$ we have the following expansion of the harmonic functions
\begin{equation}
G_i \sim \ds\frac{c_i}{\rho_i} ~, \qquad\qquad H_i \sim \ds\frac{1}{2 c_i\rho_i} ~. 
\end{equation}
Since we are looking for a regular bubbled solution in five dimensions we will assume that all functions in the solution have the same singular points (excluding the point $\rho=0$ which, as discussed in the previous section, will be treated separately).   The functions $Z_1$ and $Z_2$ near a singular point, $\rho_i \to 0$, diverge as
\begin{equation}
Z_1 \sim  \ds\frac{\ell_1^i}{c_i\rho_i} ~, \qquad\qquad Z_2 \sim  \ds\frac{\ell_2^i}{c_i\rho_i} ~. 
\label{ZIsing}
\end{equation}
To ensure regularity we should set
\begin{equation}
\ell_1^i=\ell_2^i = 0~, \qquad \forall ~i ~. 
\label{ellzero1}
\end{equation}
The function $Z_3$ near a singular point, $\rho_i \to 0$, is
\begin{equation}
Z_3 \sim  \left(\ell_3^i c_i + \ds\frac{k_1^ik_2^i}{4c_i q_i}\right)\ds\frac{1}{\rho_i} + \ds\frac{m_i}{c_i \rho_i}\left(\varepsilon_0+ \ds\sum_{k=1,k\neq i}^{N} q_k \text{sign}(c_k^2-c_i^2)\right) + \ds\frac{q_i m_i (\eta^2 -\zeta^2 + c_i^2)}{2 c_i^2 \rho_i^2}~. 
\end{equation}
The last term in the expression above is divergent and could be made to vanish only for $m_i=0$. Therefore for a regular $Z_3$ one should set
\begin{equation}
m_i= 0~,\qquad\qquad \ell_3^i = -\ds\frac{k_1^ik_2^i}{4 c_i^2 q_i}~, \qquad \forall ~i ~. 
\end{equation}
It is not hard to show that with this choice of constants the function $\mu$ will limit to a constant near a singular point. The condition for absence of CTC's\footnote{This comes from $\mathcal{Q} \geq 0$, where $\mathcal{Q}$ is defined in Appendix \ref{causality}.} requires that $\mu$ should vanish at a singular point of $V$ and this leads to the constraint:
\begin{equation}
m_0 + \ds\frac{\gamma}{ c_i^2} - \ds\frac{k_1^ik_2^i}{8 c_i^2 q_i^2} = 0~, \qquad \forall ~i ~. \label{muregrhoi}
\end{equation}
There is one more condition on the constant parameters of the solutions coming from removing a possible Dirac-Misner string in $\omega$. However it turns out that after setting $\omega_0=0$ the absence of a Dirac-Misner string in $\omega$ is guaranteed by \eqref{muregrhoi}. 

To summarize, the conditions for regularity and absence of CTC's and Dirac-Misner strings near the poles of the harmonic functions requires that we set:
\begin{equation}
m_i=\ell_1^i=\ell_2^i = 0~, \qquad \ell_3^i = -\ds\frac{k_1^ik_2^i}{4 c_i^2 q_i}~,\qquad m_0 + \ds\frac{\gamma}{ c_i^2} - \ds\frac{k_1^ik_2^i}{8 c_i^2 q_i^2} = 0  ~,\qquad \forall ~i ~. \label{singconstr}
\end{equation}
Note that these conditions are quite different from the regularity and causality constraints for BPS bubbled solutions with a GH base \cite{Bena:2007kg}. In particular for the class of bubbled solutions discussed here there is no analogue of the ``bubble equations" (or integrability conditions) familiar from the supersymmetric multi-center solutions \cite{Bena:2007kg, Bates:2003vx}. However we still have an equation that fixes the locations of the poles in the harmonic functions (but not the distance between them) in terms of the parameters $\{\gamma,m_0,k_1^i,k_2^i,q_i\}$.

Our analysis so far does not guarantee the regularity of the supergravity scalars ({\it i.e.} the K\"ahler moduli of the tori in M-theory) and the absence of causal pathologies at asymptotic infinity. To ensure that we should study the behavior of the solution at $\rho \to \infty$. The harmonic functions have the following expansion
\begin{equation}
G_i \sim 2c_i^2 \ds\frac{\zeta^2}{\rho^4}~, \qquad\qquad H_i \sim \ds\frac{1}{\rho^2} ~. 
\end{equation}
Imposing the regularity and causality constraints at $\rho\to \infty$ one finds the following constraints on the parameters of the solution: 
\begin{equation}
m_0=k_1^0k_2^0=0~, \qquad \ell_3^0- (k_1^0\ell_1^0+k_2^0\ell_2^0)=0~,\qquad k_1^0\beta_2 + k_2^0\beta_1 + k_1^0\ds\sum_{i=1}^{N}k_2^i + k_2^0\ds\sum_{i=1}^{N}k_1^i =0~. \label{asympconstr}
\end{equation}
The constraints are easily solved by imposing $\ell_3^0=k_1^0=k_2^0=m_0=0$, however there are in principle other ways to satisfy the relations in \eqref{asympconstr}, so we will not commit to a specific solution. 

The asymptotic expansion ($\rho\to\infty$) of the metric functions in the solution is
\begin{eqnarray}
Z_1 &\sim& \frac{1}{\varepsilon_0} \Big(\beta_2 +  \sum_{i=1}^{N} k_2^i \Big)  \frac{1}{\rho^2}~,  \qquad   Z_2 ~\sim~ \frac{1}{\varepsilon_0} \Big(\beta_1  +  \sum_{i=1}^{N} k_1^i \Big) \frac{1}{\rho^2}~, \notag\\
Z_3 &\sim& 2(\varepsilon_0 - Q)  \frac{\gamma}{\rho^2} +  \frac{1}{\varepsilon_0} \Big(\beta_3\varepsilon_0 + \beta_1\beta_2 + \sum_{i=1}^{N} \Big(\beta_2k_1^i+\beta_1k_2^i -  \varepsilon_0 \frac{k_1^i k_2^i}{2 q_i} \Big) +  \sum_{i,j = 1}^{N} k_1^i k_2^j   -4\varepsilon_0\gamma Q \Big)\,\frac{\zeta^2}{\rho^4}~. \notag\\
\mu &\sim& - \frac{1}{2\varepsilon_0^2} \Big(\beta_3\varepsilon_0 + 2\beta_1\beta_2 + 2 \sum_{i=1}^{N}(\beta_2k_1^i+\beta_1k_2^i)+ 2 \sum_{i,j = 1}^{N} k_1^i k_2^j - \varepsilon_0\sum_{i=1}^{N} \frac{k_1^i k_2^i}{2 q_i} -4\varepsilon_0\gamma Q \Big)  \frac{\zeta^2}{\rho^4}~.\notag
\end{eqnarray}
The constraints \eqref{asympconstr} together with \eqref{singconstr} lead to
\begin{equation}
\omega =  \frac{\beta_3}{2} \frac{\sin^2 \theta}{\rho^2} d\phi \,,
\end{equation}
It is clear that at $\rho \to \infty$ these regular bubbled solutions are asymptotic to the warped, rotating $AdS_2\times S^3$ solution presented in Section  \ref{flatsol1}. The parameters of the solution can be arranged such that the warp factor in the metric is a constant and the solution is asymptotic to the near horizon BMPV black hole.

The axisymmetric multi-center solutions have $8N+11$ parameters. The regularity and causality constraints studied in this section impose $5N+4$ relations on them, therefore we have a $(3N+7)$-parameter family of regular solutions with non-trivial topology on the base. It should be emphasized that we have only analyzed in detail the condition for absence of CTC's near the singularity of the harmonic functions and at asymptotic infinity. In principle one needs to ensure that there are no CTC's globally and for this one usually has to rely on numerics \cite{Bena:2006kb}.  On the other hand, experience with many examples suggests that once one has addressed this at singular points and ensured that the metric coefficients are well-behaved then there are no CTC's globally.

It is interesting to note that there is no analog of the bubble  equations \cite{Bena:2007kg, Bates:2003vx} for our regular non-BPS solutions.  Bubble equations can be viewed as a form of angular momentum balance that constrains the location of sources and with pure flux solutions, non-trivial bubble equations  require non-zero sources for all three fluxes.   In our solutions,  the magnetic flux of $\Theta^{(3)}$ is trivial on the topological two-cycles and the complete Maxwell field $\cF$, has no localized sources.  Thus one should not be too surprised at the absence of constraints on the location of the remaining flux sources.

\section{Conclusions}
\label{Conclusions}

Using the floating brane Ansatz of \cite{Bena:2009fi} we have constructed a large class of non-BPS multi-centered supergravity solutions. The solutions are determined by a four-dimensional K\"ahler base with non-trivial topology and that is a solution of the Euclidean Einstein-Maxwell equations. To find explicit solutions one has to solve a coupled linear system of inhomogeneous differential equations on this base. We managed to construct the most general explicit solution of these equations on the axisymmetric LeBrun-Burns base. The generic multi-centered solutions will have horizons but we showed explicitly that by a judicious choice of parameters one can make the solutions completely smooth and regular.  Due to the Maxwell flux on the four-dimensional base the five-dimensional solutions are not asymptotically flat but can be arranged to look like a warped, rotating $AdS_2\times S^3$ space at asymptotic infinity. For specific choice of parameters the asymptotic metric is exactly the near horizon throat metric of the BMPV black hole. We have thus constructed ``hair in the back of a throat".

There are a number of possible directions for further work in this area.  First, it  well-known that BPS supertubes with two electric and one magnetic dipole charge are regular in six dimensions in the D1-D5 duality frame \cite{Lunin:2001jy,Lunin:2002iz,Bena:2008dw}.  Such solutions can thus potentially provide richer classes of regular geometries.  Indeed, five-dimensional regularity requires that all the $Z_I$ be non-singular but supertubes allow two of the $Z_I$ to have poles and the singularities are resolved as Kaluza-Klein monopoles in six dimensions.  The solutions presented in Section \ref{GenAxi}, before five-dimensional regularity was imposed, include solutions that correspond to families of concentric supertubes. Removing the singularities as in  (\ref{ZIsing}) required us to set some of the parameters to zero (see (\ref{ellzero1})) and while we still found regular solutions with microstate structure,   it restricted that family of solutions quite strongly and led us to solutions for which the bubble equations were trivial.    We expect that for solutions with supertubes there will be some analog of the familiar radius formula arising from the bubble equations, or integrability conditions.   We therefore expect there to be even richer classes of bubbles and ``hair'' if one allows solutions that are regular in six dimensions but not necessarily in five.

It is also worth recalling that there are spectral flow methods that map regular, six-dimensional supertube geometries onto five-dimensional, regular bubbled geometries \cite{Bena:2008wt}.  For BPS solutions, these transformations do not substantially modify the geometry of the four-dimensional base, though they can modify the asymptotics at infinity.  On the other hand, for non-BPS solutions such spectral flows can completely change the geometry of the base, for example, mapping a hyper-K\"ahler geometry onto an Israel-Wilson electrovac solution \cite{Bena:2009fi}.  It would be interesting to see how such spectral flows might modify the solutions considered here, particularly if one first includes supertube configurations.    It will almost certainly move one beyond the LeBrun class of solutions  and perhaps give a richer  class of geometries at infinity.  

There are other natural generalizations of the solutions considered here.  Our solutions can be uplifted to eleven dimensions where they are sourced by intersecting M2 and M5 branes on $T^6$ \cite{Bena:2007kg}.   It is fairly evident that there will also be solutions that can be obtained from intersecting M2 and M5 branes wrapping two-cycles and four-cycles in a more general Calabi-Yau three-fold.  Going in the opposite direction, any solution with a LeBrun base has a space-like Killing vector (defined by $\tau$-translations) and so one can perform a dimensional reduction along this direction to find supergravity solutions in four dimensions. These solutions will clearly be non-BPS and will represent an infinite class of multi-center four-dimensional solutions that are non-supersymmetric generalizations of the solutions of \cite{Bates:2003vx}. 

It would be interesting to explore the attractor mechanism for our solutions and make connections with recent discussions on non-BPS attractors. The multi-centered solutions of Section \ref{BlowingBubbles} may realize non-BPS split attractors. It is interesting to note that in a recent discussion on non-supersymmetric split attractor flows the authors of \cite{Nampuri:2010um} also found that there are no bubble equations (or integrability conditions). This fits with our analysis in Section \ref{BlowingBubbles} and it will be very interesting to make this connection more precise.
 
Since our solutions are asymptotic to anti-de Sitter space one can do holographic analysis of the ``hair'' corresponding to our geometries and understand them as duals to states (or thermal ensembles) in the corresponding CFT.   The solutions presented here have a warped and rotating $AdS_2$ region and while the $AdS_2/CFT_1$ correspondence is not understood in such detail as its higher dimensional analogs\footnote{For a recent discussion on holography for backgrounds with and $AdS_2$ factor see \cite{Sen:2011cn}.}   there might be some effective approach similar to the one in \cite{Guica:2008mu}. 
Alternatively,  one might use a series of dualities and transform the solutions to the D1-D5-P IIB duality frame \cite{Bena:2008dw} and study the states in the D1-D5 CFT.  One might then be able to study the stability of the solutions and make some connection with the recent discussion of Hawking radiation from non-supersymmetric solutions of the D1-D5 system \cite{Chowdhury:2007jx,Avery:2009tu,Avery:2010hs}.

One would also very much like to find explicit non-supersymmetric solutions that have a throat region that looks like the solutions discussed in this paper but are asymptotically flat at infinity. To achieve this, one will probably have to find a way of breaking the relationship between the background electromagnetic field and the K\"ahler form.  To achieve this one will probably have to relax some of the simplifying assumptions of the floating brane Ansatz \cite{Bena:2009fi} and work with more general (and complicated) equations of motion.  However,  there are almost certainly even broader classes of non-supersymmetric solutions that are determined by linear systems of equations and thus such explicit non-BPS solutions may well be within reach.

\bigskip
\bigskip
\leftline{\bf Acknowledgements}
\smallskip
We would like to thank Chris Beem, Iosif Bena and Cl\'ement Ruef for useful conversations. NB would like to dedicate this work to the memory of his grandparents Kunka and Nedelcho. NB is grateful to the USC Department of Physics and Astronomy for warm hospitality while part of this work was completed. This work was supported in part by DOE grant DE-FG03-84ER-40168.  



\appendix
\section{More details on the solutions}

\renewcommand{\theequation}{A.\arabic{equation}}
\renewcommand{\thetable}{A.\arabic{table}}
\setcounter{equation}{0}
\label{appendixA}

\subsection{Gauge invariance}
\label{gauge invariance}

The general solutions on a LeBrun-Burns base discussed in Section 4 have a ``gauge invariance" similar to the one present in multi-centered BPS solutions with a GH base (see equation (94) in \cite{Bena:2007kg})). It is easy to check that the following transformation leaves \eqref{BurnsZforms} and \eqref{Burnsmuform} invariant 
\begin{eqnarray}
K^{(1)} & \to& K^{(1)} + \gamma_1 V~, \qquad\qquad K^{(2)} \to K^{(2)} + \gamma_2 V~, \\
L_1 & \to& L_1 - \gamma_2 ~, \qquad\qquad L_2 \to L_2 - \gamma_1 ~, \\
L_3 &\to& L_3 - \gamma_1 \zeta^2 K^{(2)} - \gamma_2 \zeta^2 K^{(1)} -  \gamma_1\gamma_2 \zeta^2 V~, \\
M &\to& M + \coeff{1}{2} \gamma_1 \zeta^2 L_1 + \coeff{1}{2} \gamma_2 \zeta^2 L_2  -  \coeff{1}{2} \gamma_1 \gamma_2 \zeta^2 ~,
\end{eqnarray}
One can also show that the equation for the one-form $\omega$, \eqref{Burnsomega}, is invariant, therefore the transformation above is a symmetry of the full solution. 

\subsection{Causality}
\label{causality}

A supergravity background is causal only if there are no CTCs and Dirac-Misner strings. To study the constraints imposed by these conditions one should study the five-dimensional metric at a constant time slice:
\begin{equation}
ds^2 = \mathcal{Q} \left(d\tau +A - \ds\frac{\mu V^2}{\mathcal{Q}} \omega\right)^2 + W^2 V \left( \eta^2 d\phi^2 - \ds\frac{\zeta^2}{\mathcal{Q}} \omega^2 \right) + W^2 V (d\eta^2 + d\zeta^2)~, 
\end{equation}
where
\begin{equation}
\mathcal{Q} \equiv W^6 \zeta^2 V - \mu^2 V^2~, \qquad\qquad W^2 \equiv (Z_1Z_2Z_3)^{1/3}~.
\end{equation}
For absence of CTC's we need to impose the following conditions
\begin{equation}
\mathcal{Q} \geq 0 ~, \qquad\qquad W^2V \geq 0~, \qquad\qquad Z_I^{-1} W^2 \geq 0, \qquad I=1,2,3~.
\end{equation}
The last inequality comes from imposing positive definite metric in the six internal directions along $T^6$ upon uplift of our solutions to eleven-dimensional supergavity. The expression for $\mathcal{Q}$ resembles quite closely the one for solutions with a GH base (see equation (102) in \cite{Bena:2007kg})
\begin{multline}
\mathcal{Q} = - M^2 V^2 + 2 M\zeta^2 K^{(1)} K^{(2)} + MV \left(\zeta^2 K^{(1)} L_1 + \zeta^2 K^{(2)} L_2 + L_3\right) -\coeff{1}{4} \left(\zeta^2 K^{(1)} L_1 + \zeta^2 K^{(2)} L_2 + L_3\right)^2 \\
+ \zeta^2 V L_1L_2L_3 + \left(\zeta^4 K^{(1)} L_1 K^{(2)}L_2 + L_3\zeta^2 K^{(1)} L_1 + L_3 \zeta^2 K^{(2)}L_2\right)
\end{multline}
There is also the possibility of having Dirac-Misner strings in $\omega$. To ensure that this does not happen one has to require that $\omega_{\phi}$ vanishes for $\eta=0$.

\subsection{Useful identities}
\label{UseIdents}

Here we collect some identities used in Section \ref{BlowingBubbles}. We used the following identities to solve the equations for $K^{(a)}$ and $M$ 
\begin{eqnarray}
\mathcal{L}_1\left( \ds\frac{1}{\zeta^2} \right) &=& \ds\frac{4}{\zeta^4}~, \notag\\
\mathcal{L}_1\left(- \ds\frac{1+G_i}{\zeta^2} \right) &=& \ds\frac{2}{\zeta} \partial_{\zeta} \left( \ds\frac{1+G_i}{\zeta^2} \right)~, \\ 
\mathcal{L}_1\left(- \ds\frac{(1+G_i)(1+ G_j)}{\zeta^2} + 4 \rho^2 H_i H_j \right) &=& \ds\frac{2}{\zeta} \partial_{\zeta} \left( \ds\frac{(1+G_i)(1+ G_j)}{\zeta^2} \right)~. \notag
\end{eqnarray}
The following identities are useful when one solves the equation for $L_3$
\begin{equation}
\mathcal{L}_2\left( \ds\frac{1}{\zeta^2} \right) = \ds\frac{8}{\zeta^4}~, \qquad \mathcal{L}_2\left(- \ds\frac{1+G_i}{\zeta^2} \right) = \ds\frac{4}{\zeta} \partial_{\zeta} \left( \ds\frac{1+G_i}{\zeta^2} \right)~, \qquad \mathcal{L}_2\left(- H_i \right) = \ds\frac{2}{\zeta} \partial_{\zeta} H_i~,
\end{equation}
\begin{eqnarray}
\mathcal{L}_2\left(- \ds\frac{(1+G_i)(1+ G_j)}{\zeta^2} + 4 \rho^2 H_i H_j \right) &=& \ds\frac{4}{\zeta} \partial_{\zeta} \left( \ds\frac{(1+G_i)(1+ G_j)}{\zeta^2} \right) - \ds\frac{8}{\zeta}\partial_{\zeta}(\rho^2 H_iH_j)~,\notag\\
\mathcal{L}_2\left(- \ds\frac{1}{c_i^2-c_j^2}\ds\frac{H_j - H_i}{H_i} \right) &=& \ds\frac{2}{\zeta} (1+G_i) \partial_{\zeta} H_j~,\qquad i\neq j\\
\mathcal{L}_2\left(- (\rho^2+c_i^2 - 2 \zeta^2 )H_i^2 \right) &=& \ds\frac{2}{\zeta} (1+G_i) \partial_{\zeta} H_i~, \notag
\end{eqnarray}
\begin{multline}
\mathcal{L}_2\left(- \ds\frac{(1+G_i)(1+ G_j)(1+G_k)}{\zeta^2} + 4 \rho^2 (3\rho^2 - 4\zeta^2+ c_i^2+c_j^2+c_k^2) H_i H_j H_k\right) \\ = \ds\frac{4}{\zeta} \partial_{\zeta} \left( \ds\frac{(1+G_i)(1+ G_j)(1+G_k)}{\zeta^2} \right) - \ds\frac{8}{\zeta} \Big[(1+G_i)\partial_{\zeta}(\rho^2 H_jH_k) \\ +(1+G_j)\partial_{\zeta}(\rho^2 H_kH_i)+(1+G_k)\partial_{\zeta}(\rho^2 H_iH_j) \Big]~. 
\end{multline}
There are similar identities involving $D_i$, $H_i$ and $G_i$ that we have used to solve the equation for $\omega_{\phi}$, however they are pretty lengthy and we refrain from presenting them explicitly.

\subsection{Black ring coordinates}
\label{CanCoords}

To facilitate comparison of our solutions with the more standard black-ring and supertube solutions, it is useful to recall the canonical separable bipolar coordinates  on $\IR^4$, \cite{Emparan:2001wn} (we set $a=b=0$ below):  
\begin{eqnarray}
\tilde x &\equiv& -(G+1- 2 c^2  H) ~=~ -{x^2+ y^2 + \zeta^2 - c^2 \over  \sqrt{((\zeta-c)^2 + x^2+ y^2)( (\zeta +c)^2 +x^2+ y^2  )}} \,, \\
\tilde y & \equiv& -(G+1)~=~   -{ x^2+ y^2 + \zeta^2 + c^2 \over  \sqrt{((\zeta-c)^2 + x^2+ y^2)( (\zeta +c)^2 +x^2+ y^2  )}} \,,
\end{eqnarray}
In these coordinates the flat metric on $\IR^4$ takes the form:
\begin{equation}
ds^2_{\IR^4}= {R^2 \over (\tilde x-\tilde y)^2}\left( {d \tilde y^2 \over \tilde y^2 -1} 
+ (\tilde y^2-1) d\tau^2  +{d \tilde x^2 \over 1-\tilde x^2} + (1-\tilde x^2) d \phi^2  \right)\,.
\end{equation}
where $R=c$. In particular, note that the canonical coordinates, $\tilde x$ and $\tilde y$, are simply related to the Green functions that we have been using and thus the solutions of Sections 5 and 6 can easily be expressed as rational functions of $\tilde x$ and $\tilde y$.




\end{document}